\documentclass[
reprint,
groupedaddress,
amsmath,
amssymb,
aps,
prl,
floatfix,
nofootinbib
]{revtex4-2}
\usepackage[T1]{fontenc}           
\usepackage{newtxtext,newtxmath}   
\usepackage{microtype}             

\usepackage{graphicx}
\usepackage{dcolumn}
\usepackage{bm}
\usepackage{xcolor}
\usepackage{stackengine}
\usepackage[percent]{overpic}
\usepackage{physics}
\usepackage{booktabs}
\newcommand{\Ires}{\mathcal{I}_{\mathrm{res}}}

\usepackage{tikz}
\usetikzlibrary{fpu}
\usetikzlibrary{arrows.meta,calc,decorations.pathmorphing,positioning}

\definecolor{SpinBlue}{RGB}{31,100,180}       
\definecolor{SpinRed}{RGB}{192,40,40}         
\definecolor{BathGray}{RGB}{120,130,140}      
\definecolor{JViolet}{RGB}{120,50,160}        
\definecolor{QAmber}{RGB}{200,120,0}          
\definecolor{FlipViolet}{RGB}{110,45,155}     
\definecolor{CurveRed}{RGB}{200,50,50}        
\definecolor{CurveBlue}{RGB}{30,100,185}      
\definecolor{ResGold}{RGB}{255,200,50}        
\definecolor{prlred}{HTML}{A33A3A}
\definecolor{prldarkred}{HTML}{7F2E2E}
\definecolor{prlgray}{HTML}{6F6F6F}
\definecolor{prllightgray}{HTML}{D9D9D9}

\makeatletter
\g@addto@macro\normalsize{%
	\setlength\abovedisplayskip{4pt plus 1pt minus 1pt}%
	\setlength\belowdisplayskip{4pt plus 1pt minus 1pt}%
	\setlength\abovedisplayshortskip{2pt plus 1pt}%
	\setlength\belowdisplayshortskip{2pt plus 1pt}%
}
\makeatother

\usepackage{hyperref}
\hypersetup{
	colorlinks=true,
	linkcolor=blue,
	citecolor=blue,
	urlcolor=blue
}
\begin{document}
	
	\title{Ising Blockade of Resonant Transport in Dense Spin Ensembles}
	
	\author{Hong-Ze Ding}
	\email{dinghongze@sxu.edu.cn}
	
	\author{Jiu-Qing Liang}
	\email{jqliang@sxu.edu.cn}
	
	\affiliation{
		Institute of Theoretical Physics,
		Shanxi University,
		Taiyuan 030006, China
	}
\begin{abstract}
	Dense ensembles of nitrogen-vacancy (NV), substitutional-nitrogen (P1), and surface spins in diamond are leading platforms for quantum sensing, masers, and many-body physics, and in all of them polarization moves by resonant flip-flops (pairwise spin exchanges). We show that the Ising part of the dipolar Hamiltonian, which moves no polarization, sets how fast polarization moves. It detunes every flip-flop by exactly the difference of the Ising fields that the other spins exert on the pair, and these fields come from the very couplings that drive hopping. Dipolar transport thus throttles its own rate, an effect we call the Ising blockade, which models have so far absorbed into a measured or fitted disorder width. Deep in the blockade and at fixed composition, the diffusion constant scales as $D\propto J_0\rho^{1/3}$ (coupling $J_0$, density $\rho$) instead of the F\"orster-type $J_0^2\rho^{4/3}$. Without fitting, the theory predicts the NV linewidth and all five P1 diffusion constants of [\href{https://doi.org/10.1038/s41586-021-03763-1}{Nature \textbf{597}, 45 (2021)}] within $2\sigma$; without the Ising term at the same couplings, it overshoots them five- to fifteenfold. On a spin-locked surface [\href{https://doi.org/10.1103/PhysRevLett.134.050801}{Phys.\ Rev.\ Lett.\ \textbf{134}, 050801 (2025)}], the same blockade acts coherently: in exact dynamics, over a third of the spins stay polarized far longer than flip-flops alone allow, accounting for the measured four- to ninefold slowdown.
\end{abstract}
	\maketitle
	\twocolumngrid
	
	\textit{Introduction.}---Nitrogen-vacancy (NV) and substitutional-nitrogen (P1) centers in diamond and electron spins on its surface form dense dipolar spin ensembles, used for quantum sensing and entanglement-enhanced metrology~\cite{degen2017,zhou2020,wu2025,wolfowicz2021,hallDetectionNanoscaleElectron2016}, as active media in microwave quantum memories and masers~\cite{putzSpectralHoleBurning2017,angerer2018,breezeContinuouswaveRoomtemperatureDiamond2018,kersten2023,kersten2026}, and as model systems for nonequilibrium many-body physics~\cite{choi2017,kucsko2018,choi2019,zu2021,davis2023,martin2023,he2023,hughes2025,yoon2026}. Higher spin density strengthens collective coupling and improves sensitivity, so experiments keep moving to denser samples. In this regime, all of these applications depend on how polarization migrates through the disordered dipolar network by resonant flip-flops, the only part of the dipolar interaction that moves it. This migration limits sensor coherence, sets how quickly a burnt spectral hole refills~\cite{putzSpectralHoleBurning2017,kersten2026}, and is the microscopic route to thermalization~\cite{yao2014,burin2015,abanin2019}. Spin transport is thus where reversible quantum dynamics turns into irreversible relaxation, and two experiments now observe it from opposite ends, the collective and the single spin. In the three-dimensional P1 bath, Zu \textit{et al.}~\cite{zu2021} used NV centers as local sources and probes of polarization and observed spin diffusion emerge from microscopic flip-flops, hydrodynamics arising in a strongly interacting quantum many-body system. On the diamond surface, Rezai \textit{et al.}~\cite{rezai2025} used a shallow NV center as a single-qubit sensor to follow one spin of a two-dimensional ensemble as it loses its polarization to its neighbors. Under a spin lock, which suppresses on-site fields and leaves only the dipolar coupling, it relaxes on a drive-independent plateau, $T_{1\rho}\approx 14$--$28\,T_2$, more than an order of magnitude slower than the echo decay set by the same couplings. Here we show that both the diffusion and the plateau are set by the Ising interaction, at a strength that the transporting spins generate themselves and that we compute with no measured or fitted width. This interaction conserves every local population and moves no polarization, yet it sets how fast polarization moves.
	\begin{figure}
		\centering
		\begin{overpic}[width=0.625\columnwidth]{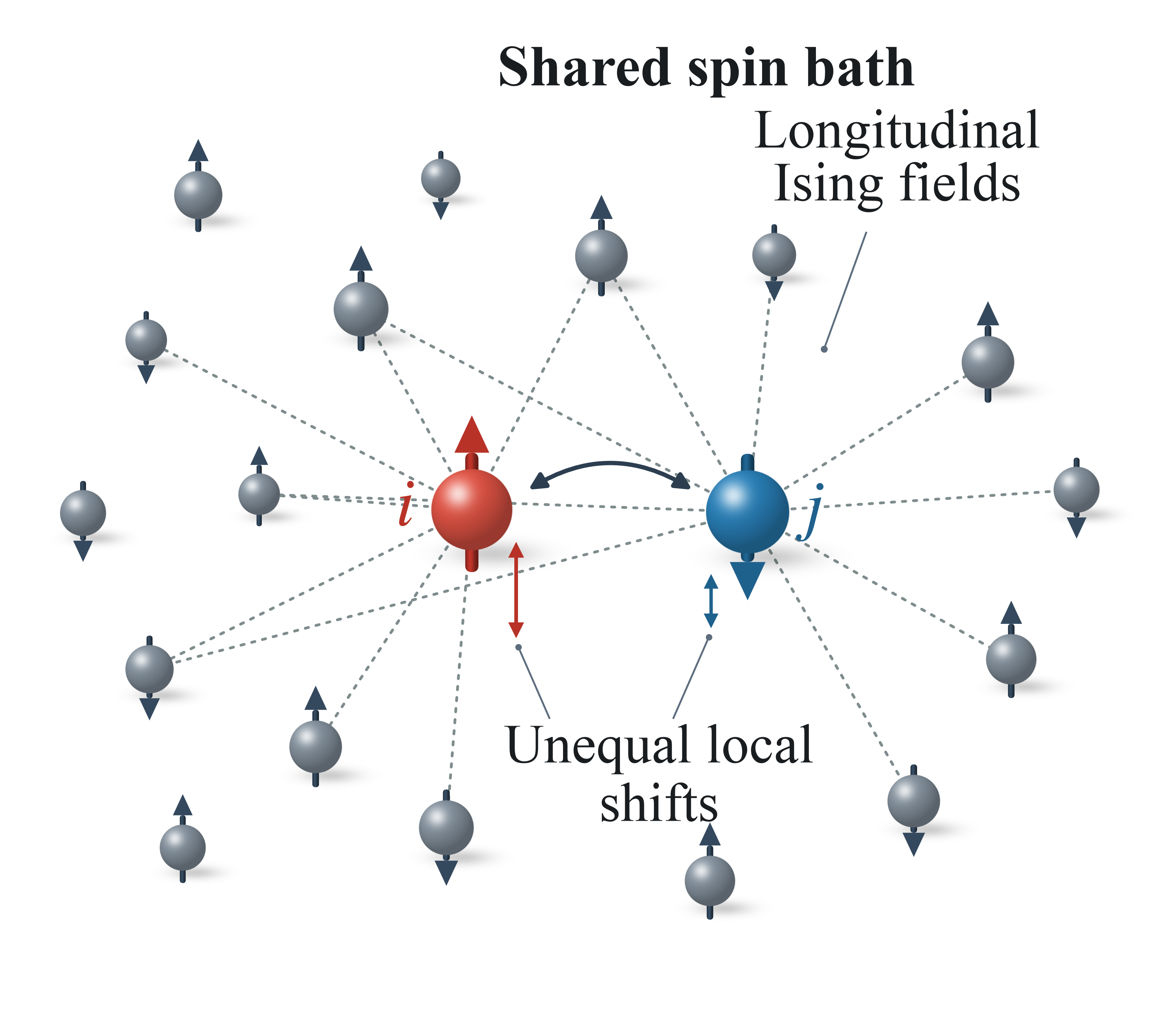}
			\put(-1,79){\makebox(0,0)[lt]{\fontsize{8.5}{10}\selectfont\bfseries (a)}}
		\end{overpic}\hfill
		\begin{overpic}[width=0.365\columnwidth]{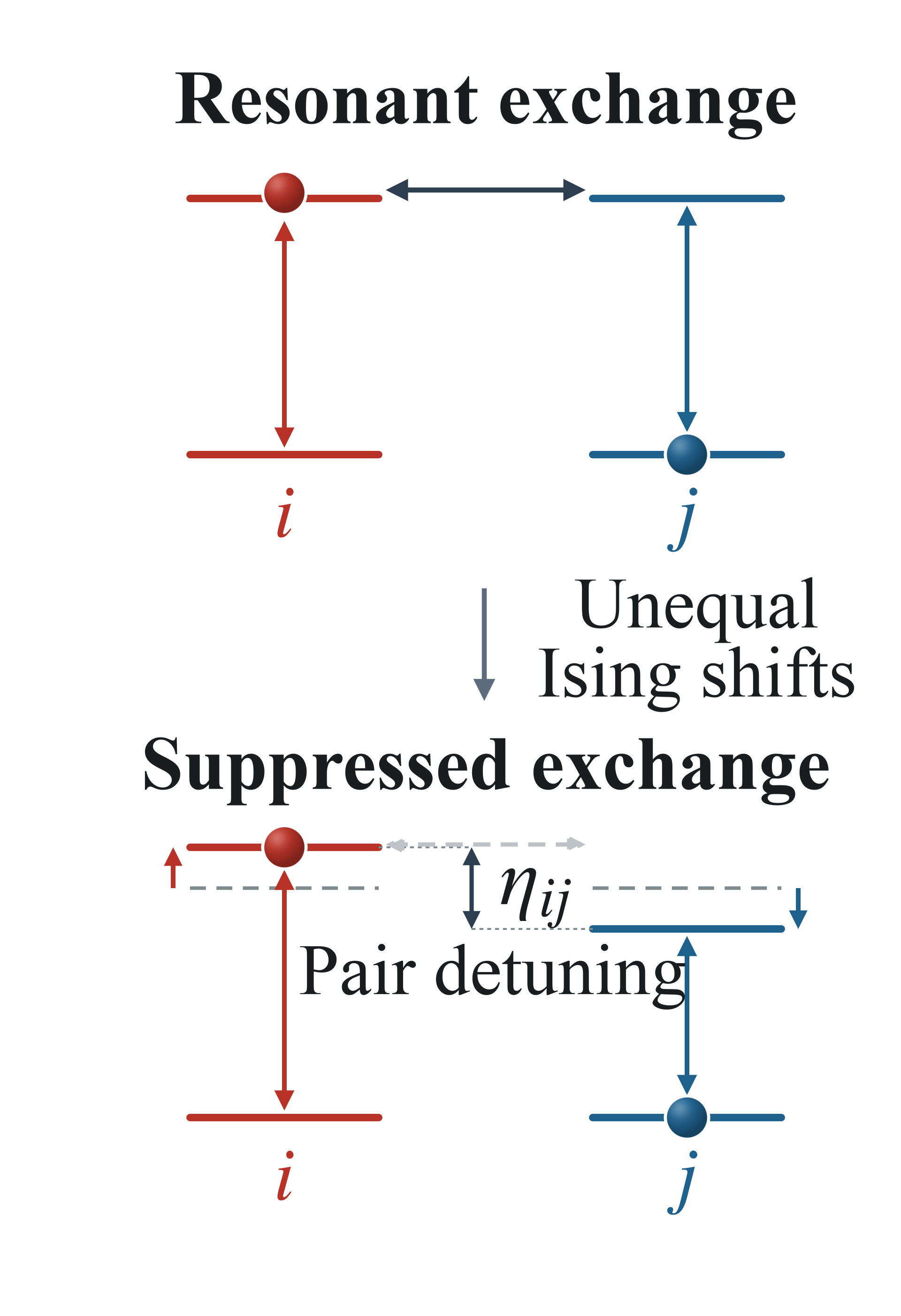}
			\put(1,92){\makebox(0,0)[lt]{\fontsize{8.5}{10}\selectfont\bfseries (b)}}
		\end{overpic}
		\caption{(Color online) \textbf{Microscopic mechanism of the Ising blockade.} 
			(a) Spins $i$ and $j$ in a shared dipolar bath feel fluctuating longitudinal Ising fields $\sum_{k\neq i,j} Q_{ik} S_k^z$ and $\sum_{k\neq i,j} Q_{jk} S_k^z$ from the other spins that induce unequal local energy shifts.
			(b) While unshifted spins undergo resonant flip-flop exchange (top), unequal Ising shifts generate an instantaneous pair detuning $\eta_{ij}$, the difference of these two fields [Eq.~\eqref{eq:eta}], that takes the pair off resonance and suppresses transport (bottom).}
		\label{fig:mechanism}
	\end{figure}
	
	In the standard theory~\cite{vanvleck1948,bloembergen1949}, Fermi's golden rule gives pairwise $r^{-6}$ flip-flop rates, evaluated with a single-spin linewidth. In a random ensemble the Van Vleck second moment of that line diverges at short distances, so the linewidth is injected from measurement~\cite{kersten2026,zu2021} or fitted~\cite{rezai2025} rather than computed from the Hamiltonian. The Ising interaction $Q_{jk}S_j^zS_k^z$, the only other secular term, transfers no excitation and has therefore been treated as static disorder inside that linewidth~\cite{vanvleck1948,abragam2011}. The two experiments went further and found that Ising fields slow transport. Zu \textit{et al.}\ traced the on-site fields mainly to Ising couplings between P1 groups and sped up diffusion by driving one group~\cite{zu2021}. Rezai \textit{et al.}, following the resonance counting of Ref.~\cite{kucsko2018}, attributed the plateau to Ising disorder that blocks flip-flops between spins whose local shifts differ by more than their coupling~\cite{rezai2025}. In both cases the width of this disorder remains an input, and the resonance-counting model of Ref.~\cite{rezai2025} has no finite solution in the plateau~\cite{SM}. Disorder that the transporting spins generate themselves is thus handled as if it came from outside.
	
	Two spins that exchange an excitation sit at different positions in the same bath and feel different longitudinal Ising fields [Fig.~\ref{fig:mechanism}(a)]~\cite{yao2014,burin2015,anderson1958}. A flip-flop is resonant only when the difference of these shifts, the pair detuning $\eta_{ij}$, is small compared with the homogeneous linewidth [Fig.~\ref{fig:mechanism}(b)]~\cite{klauder1962,anderson1953,blackSpectralDiffusionPhonon1977}. This shift is exact: a flip-flop changes the Ising energy by exactly $\eta_{ij}$ and leaves its matrix element unchanged. The Ising term therefore enters transport only through the weight of the $\eta_{ij}$ distribution near zero detuning, and this distribution is generated by the very couplings that drive hopping. Raising the coupling or the density speeds up hopping but spreads $\eta_{ij}$ and leaves fewer pairs resonant: dipolar transport limits its own rate. By analogy with blockade effects in Rydberg ensembles~\cite{lukin2001} and trapped-ion magnets~\cite{bermudez2016}, we call this suppression the Ising blockade. The mechanism requires only an Ising coupling that varies from spin to spin, so it applies to disordered XXZ models in general.
	
\begin{figure}
		\centering
		\includegraphics[width=\columnwidth]{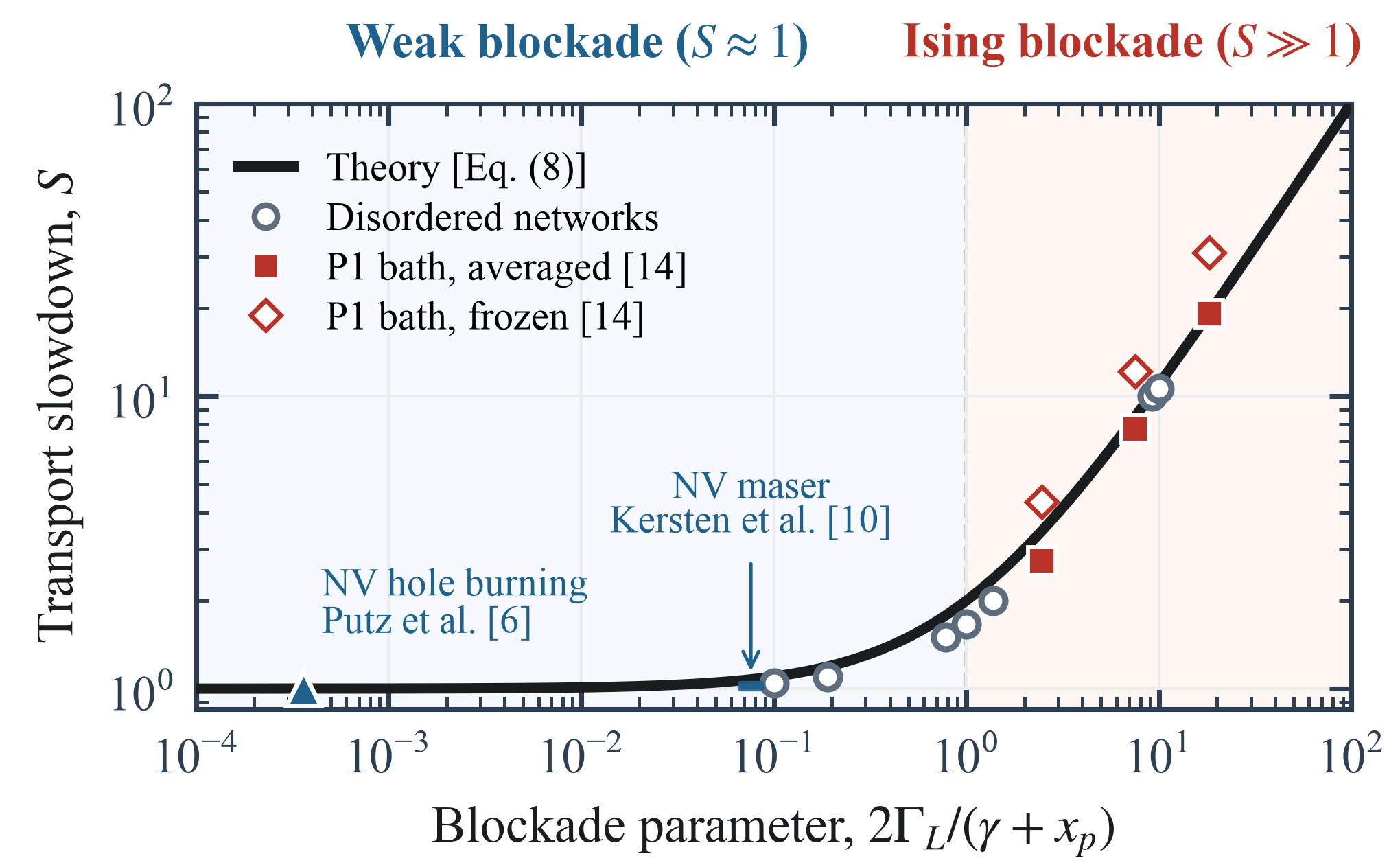}
		\caption{(Color online) \textbf{Universal platform crossover and the Ising blockade criterion.} 
			Transport slowdown $S$, the ratio of $D$ without and with the Ising term at fixed $J_{ij}$, as a function of the parameter-free ratio $2\Gamma_L/(\gamma + x_p)$, with $\Gamma_L$ the Ising half width [Eq.~\eqref{eq:GammaL}] and $\gamma+x_p$ the non-Ising pair width.
			Black solid curve is the universal analytical prediction $S_\infty = 1 + 2\Gamma_L/(\gamma + x_p)$ [Eq.~\eqref{eq:criterion}].
			Filled squares and open diamonds: networks at the parameters of Ref.~\cite{zu2021} with bath-averaged and frozen-bath rates.
			Dense NV ensembles~\cite{putzSpectralHoleBurning2017,kersten2026} operate in the disorder-dominated weak-blockade regime ($S \approx 1$), whereas the diamond P1 bath~\cite{zu2021} lies deep in the self-throttled Ising blockade ($S_\infty \simeq 2\Gamma_L/(\gamma+x_p)$).}
		\label{fig:crossover}
	\end{figure}

	In spectral-diffusion theory the fluctuating longitudinal field dephases a spin~\cite{klauder1962,blackSpectralDiffusionPhonon1977,desousaTheoryNuclearinducedSpectral2003}, and in resonance counting it blocks flip-flops~\cite{kucsko2018,rezai2025}; what has been missing is the distribution of $\eta_{ij}$, without which the Ising width must be taken from experiment. In this Letter, we develop a microscopic theory of the Ising blockade with no adjustable parameter. In a random $d$-dimensional ensemble, $\eta_{ij}$ obeys a universal stable law~\cite{klauder1962} fixed by the $r^{-3}$ coupling and the dimension alone, with no short-distance cutoff. Because the flip-flop and Ising couplings scale identically with density, a single density-independent number decides whether the strongest bond of a spin transfers incoherently or coherently. It places the P1 bath in incoherent hopping and the surface ensemble in coherent exchange, two regimes of one blockade. Deep in the blockade, spin diffusion collapses from $D\propto J_0^2\rho^{4/3}$ to $D\propto J_0\rho^{1/3}$, where $J_0$ is the dipolar coupling coefficient and $\rho$ the spin density, so doubling the density speeds up diffusion by only 26\% instead of 150\%. Without fitting, the theory reproduces the NV probe linewidth and, within $2\sigma$, all five P1 diffusion constants of Ref.~\cite{zu2021}; without the Ising term, at identical flip-flop couplings, they would be five to fifteen times the measured values. On the surface, the Ising blockade supplies the factor of four to nine by which the spin-lock plateau exceeds the relaxation time that flip-flops alone would give. Dense NV ensembles~\cite{putzSpectralHoleBurning2017,kersten2026}, by contrast, lie in the weak-blockade regime [Fig.~\ref{fig:crossover}]. Global Hamiltonian engineering offers no escape: for like spin-$\frac12$ dipoles such as P1 centers and surface spins, no such control can remove the blockade without halting transport.

	\textit{Pair detuning.}---The secular dipole-dipole Hamiltonian ($\hbar=1$) is
	\begin{equation}
		H_{\rm dd}=\sum_{j<k}\left[J_{jk}S^+_jS^-_k+J^*_{jk}S^-_jS^+_k+Q_{jk}S^z_jS^z_k\right].
		\label{eq:H}
	\end{equation}
	For like spin-$\frac12$ centers quantized along $\mathbf B$, such as P1 centers and surface spins, $Q_{jk}=J_0a_{jk}/r_{jk}^3$ with $a_{jk}=1-3\cos^2\theta_{jk}$, $J_0/2\pi=52.0$~MHz\,nm$^3$, and $J_{jk}=-Q_{jk}/4$, so the anisotropy is $\lambda\equiv|Q_{jk}/J_{jk}|=4$ for every pair ($\lambda=2$ for NV centers in $\{|0\rangle,|{-1}\rangle\}$). Writing the Ising term as $\frac12\sum_j\delta\omega_jS^z_j$ with $\delta\omega_j=\sum_{k\neq j}Q_{jk}S^z_k$ identifies the local frequency shift of spin $j$. Two spins that exchange an excitation sit at different positions in the same bath, and the shifts induced on them by all spins other than the pair differ by the pair detuning [Fig.~\ref{fig:mechanism}(a)]
	\begin{equation}
		\eta_{ij}\equiv\sum_{k\neq i,j}\left(Q_{ik}-Q_{jk}\right)S^z_k .
		\label{eq:eta}
	\end{equation}
	No approximation is involved. A flip-flop $S^+_iS^-_j$ changes $H_Q\equiv\sum_{j<k}Q_{jk}S^z_jS^z_k$ by exactly $\eta_{ij}$, because the mutual term $Q_{ij}S^z_iS^z_j$ is the same before and after the flip-flop, so that~\cite{SM}
	\begin{equation}
		e^{iH_Qt}\,S^+_iS^-_j\,e^{-iH_Qt}=e^{i\eta_{ij}t}\,S^+_iS^-_j .
		\label{eq:identity}
	\end{equation}
	In the interaction picture with respect to $H_Q$, the dynamics is generated entirely by $\sum_{i<j}J_{ij}e^{i\eta_{ij}t}S^+_iS^-_j+{\rm H.c.}$ Without the Ising term, every flip-flop is exactly resonant for any configuration of the other spins. With it, a flip-flop is detuned, but its matrix element is unchanged. For frozen bath spins, a bond transfers a fraction $A_{ij}=4J_{ij}^2/\omega_{ij}^2$ at the Rabi frequency $\omega_{ij}=(\eta_{ij}^2+4J_{ij}^2)^{1/2}$, while $\frac12A_{ij}\omega_{ij}^2=2J_{ij}^2$ stays fixed~\cite{SM}. The Ising term thus trades transferable weight for frequency at constant second moment [Fig.~\ref{fig:mechanism}(b)].
	
	\textit{Resonant window.}---If $\eta_{ij}$ fluctuates during a transfer and its spread exceeds $J_{ij}$, the transfer is incoherent, and Fermi's golden rule gives the pair rate
	\begin{equation}
		W_{ij}=2\pi|J_{ij}|^2\,\Ires,\qquad \Ires=\int P_{ij}(\Delta)\,L_\gamma(\Delta)\,d\Delta ,
		\label{eq:FGR}
	\end{equation}
	where $P_{ij}(\Delta)$ is the distribution of the total pair detuning ($\eta_{ij}$ plus static non-Ising mismatches) and $L_\gamma$ is the homogeneous Lorentzian of half width $\gamma$. By Eq.~\eqref{eq:identity}, the Ising term enters $W_{ij}$ only through $P_{ij}$, so the resonant window $\Ires$ is the weight of the pair detuning at resonance rather than a single-spin linewidth.
	
	\textit{Universal statistics.}---Let the Ising sources be Poisson distributed with density $\rho$ in $d$ dimensions, with $S^z_k=\pm\frac12$. The single-site field $h_j=\sum_kQ_{jk}S^z_k$ then has the characteristic function~\cite{klauder1962,zolotarev1986}
	\begin{equation}
		\begin{aligned}
			\langle e^{ikh_j}\rangle&=e^{-c_h|k|^{d/3}},\\
			c_h&=K_d\,\rho\Big(\frac{J_0}{2}\Big)^{d/3}\!\!\int|a(\Omega)|^{d/3}d\Omega ,
		\end{aligned}
		\label{eq:stable}
	\end{equation}
	with $K_3=\pi/6$ and $K_2=\Gamma(\frac13)/4$ [End Matter~A]. This symmetric stable law of index $d/3$ is fixed by the $r^{-3}$ coupling and the $d$-dimensional measure alone, with neither a cutoff nor a lattice constant. The second moment that defines the Van Vleck linewidth~\cite{vanvleck1948} diverges at short distances, whereas the core that enters Eq.~\eqref{eq:FGR} stays finite. In three dimensions, Eq.~\eqref{eq:stable} is the Lorentzian of dilute dipolar solids~\cite{kittel1953,klauder1962} with half width
	\begin{equation}
		\Gamma_L=\frac{4\pi^2}{9\sqrt3}\,J_0\rho ,
		\label{eq:GammaL}
	\end{equation}
	using $\langle|a|\rangle=4/(3\sqrt3)$; in two dimensions the law has index $2/3$. For well-separated pairs the fields at the two ends are independent~\cite{SM}, so $\eta_{ij}$ follows the same law with $c_h\to2c_h$; in three dimensions it is a Lorentzian of half width $x_h=2\Gamma_L$. The pair-detuning distribution is therefore broader than the single-site one by $2^{3/d}$, i.e., by 2 in three dimensions and $2^{3/2}$ in two, compared with $\sqrt2$ for a Gaussian line.
	
	\textit{Two regimes.}---The same statistics determine whether the strongest bond of a spin is incoherent or coherent. The golden-rule rate~\eqref{eq:FGR} reaches the coherent transfer rate $2|J_{ij}|$ when $\pi|J_{ij}|P_{ij}(0)=1$. Let $J_1$ be the strongest flip-flop coupling of a spin to partners of density $\nu\rho$, and $|J_1|_{1/2}$ its median. Its distribution is known in closed form, and with Eq.~\eqref{eq:stable} it gives [End Matter~A]
	\begin{equation}
		2\pi|J_1|_{1/2}P_{ij}(0)=\frac4\lambda
		\begin{cases}
			\nu/(\pi\ln2), & d=3,\\[3pt]
			\Gamma(\frac52)\,[\Gamma(\frac13)\ln2]^{-3/2}\,\nu^{3/2}, & d=2,
		\end{cases}
		\label{eq:regime}
	\end{equation}
	i.e., $0.459\,\nu$ and $0.525\,\nu^{3/2}$ at $\lambda=4$, independent of the density and of the field direction because both factors scale with the same angular integral. In the P1 bath every center is an Ising source, but only one hyperfine group, of weight $\nu=\frac13,\frac14,\frac1{12}$, provides flip-flop partners. Equation~\eqref{eq:regime} then gives 0.15, 0.11, and 0.04, and only 1--5\% of the spins have a strongest bond beyond the golden-rule bound, so transport is incoherent hopping. On a surface with a single species, $\nu=1$ and Eq.~\eqref{eq:regime} gives 0.53, so a quarter of the spins have a coherent strongest bond. The two experiments thus probe the two regimes of the same blockade, and we treat each with the method its regime requires.
	
	\begin{figure*}[t]
		\centering
		\includegraphics[width=0.48\textwidth]{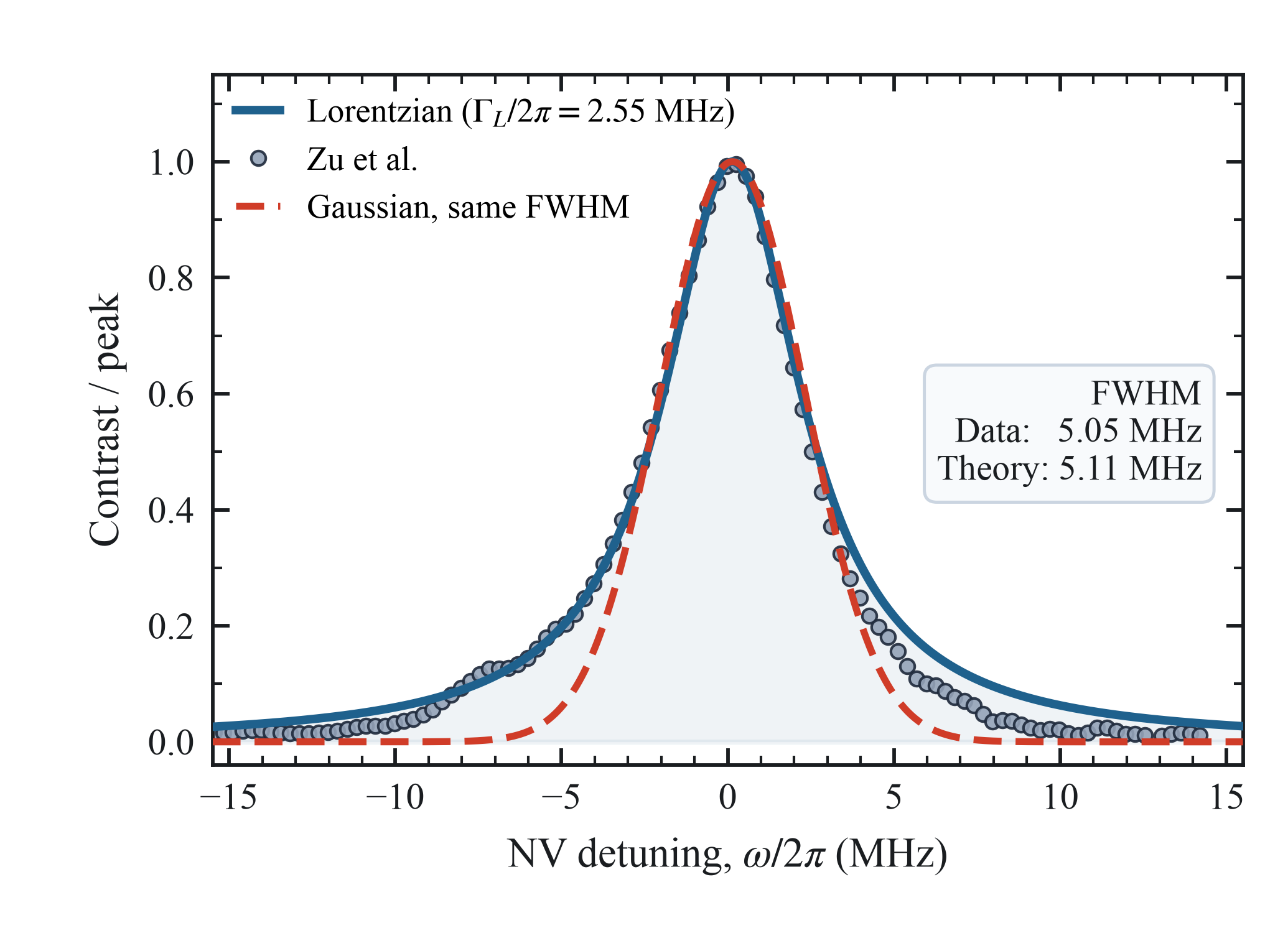}\hfill
		\includegraphics[width=0.48\textwidth]{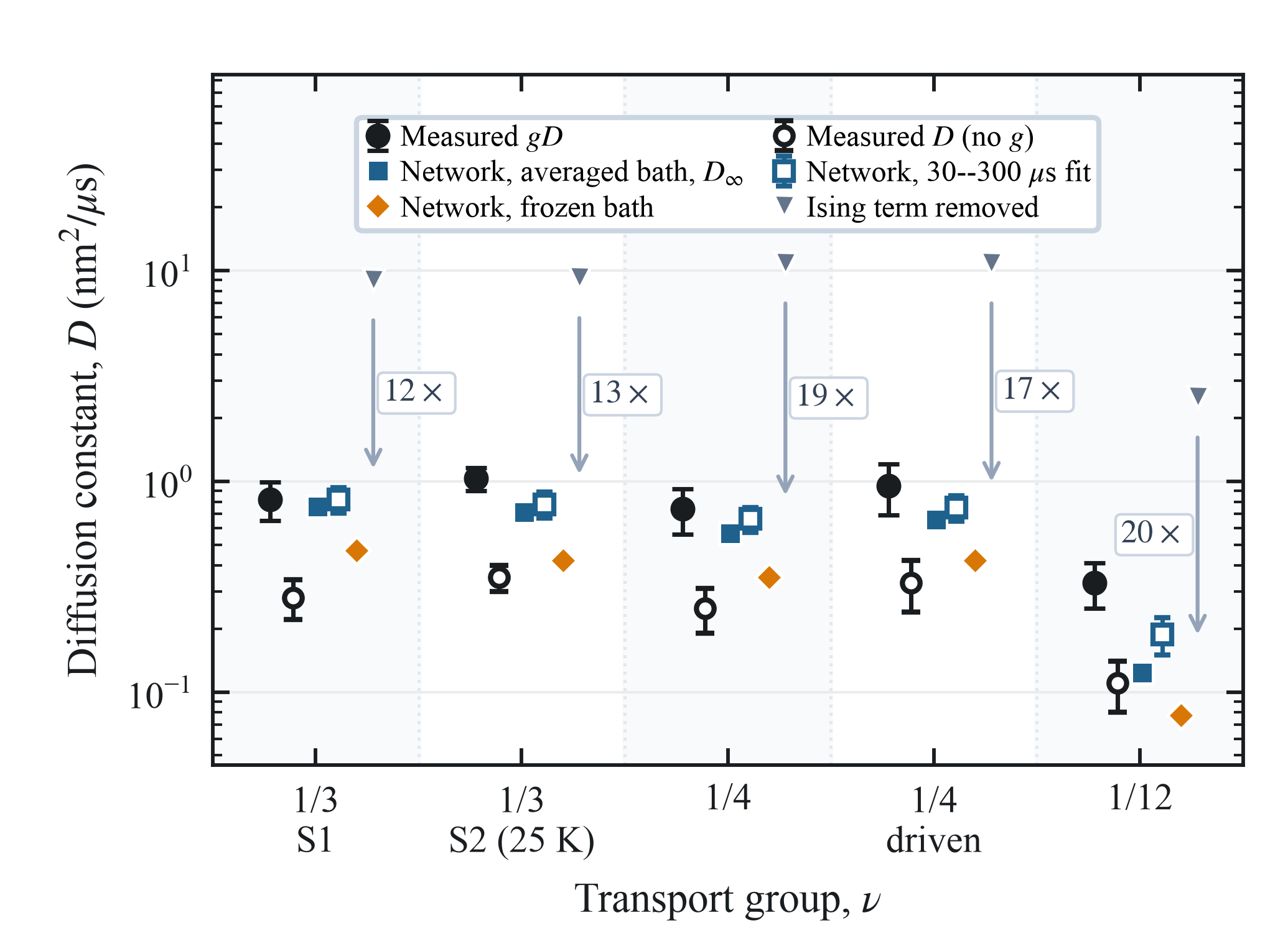}
		
		\caption{\textbf{The Ising term sets both the NV linewidth and P1 spin diffusion.}
			(a) NV spectrum measured in Ref.~\cite{zu2021} (circles), compared with the Lorentzian
			of Eq.~\eqref{eq:GammaL}, $\Gamma_L/2\pi = 2.55$~MHz, with no free parameter (solid), and with a
			Gaussian of the same full width (dashed), which leaves twice the residual.
			(b) Spin-diffusion constants of the five transport groups. Measured values of
			Ref.~\cite{zu2021}: $gD$ (filled circles) and raw $D$ (open circles), with
			$g = 2\pi^{1/3}$. Master-equation networks with all five groups as Ising sources:
			long-time $D_\infty$ for bath-averaged rates (filled squares), fits of
			$\langle r^2 \rangle$ over the experimental window of 30--300~$\mu$s (open squares),
			and a frozen bath (diamonds). Triangles: the same networks with the Ising term
			removed at fixed $J_{ij}$; arrows give their ratio to $D_\infty$, and they lie 8--15 times above the measured $gD$. Inputs as in
			Table~\ref{tab:inputs}, with no fitted parameter.}
		\label{fig:2}
	\end{figure*}
	
	\textit{Blockade criterion and spin diffusion.}---In the incoherent regime the role of the Ising term is explicit. Static non-Ising mismatches, such as hyperfine fields, add a pair width $x_p=2x_s$ for a Lorentzian single-spin line of half width $x_s$. Since Lorentzian widths add under convolution,
	\begin{equation}
		\Ires=\frac{1}{\pi(\gamma+x_p+2\Gamma_L)},\qquad S_\infty=1+\frac{2\Gamma_L}{\gamma+x_p},
		\label{eq:criterion}
	\end{equation}
	where $S_\infty$ is the factor by which the Ising term slows a pair rate at fixed $J_{ij}$. The blockade dominates when $2\Gamma_L\gtrsim\gamma+x_p$. Deep in this regime the window is set by the Ising field alone, so $W_{ij}\propto|J_{ij}|^2/J_0\rho$ and the interaction competes with itself. The populations obey the Pauli master equation $\dot n_i=\sum_jW_{ij}(n_j-n_i)$, a random walk on the network of rates~\eqref{eq:FGR} that becomes diffusion at long times~\cite{SM}. For a common window, $W_{ij}\propto a_{ij}^2/r_{ij}^6$, the only length scale is $(\nu\rho)^{-1/3}$, and
	\begin{equation}
		D=\frac{c}{8}\,\frac{J_0^2(\nu\rho)^{4/3}}{\gamma+x_p+2\Gamma_L},
		\label{eq:D}
	\end{equation}
	
	where $c=1.94\pm0.01$ is a pure number that we obtain by solving the homogenization problem on Poisson networks~\cite{SM}, the dipolar analog of the F\"orster prefactor~\cite{haan1978,gochanour1979}. Without the Ising term, $D\propto J_0^2\rho^{4/3}$. Deep in the blockade,
	\begin{equation}
		D\simeq\frac{9\sqrt3\,c}{64\pi^2}\,J_0\,\nu^{4/3}\rho^{1/3}=0.048\,J_0\,\nu^{4/3}\rho^{1/3},
		\label{eq:Ddeep}
	\end{equation}
	so doubling the density speeds up diffusion by only 26\% instead of 150\%.

	\textit{Test 1: spin diffusion in the P1 bath.}---Zu \textit{et al.}~\cite{zu2021} studied P1 centers at 110~ppm ($\rho=0.0194$~nm$^{-3}$) with $B=511$~G along an NV axis. The $^{14}$N hyperfine interaction and the four Jahn-Teller axes split the P1 line into five groups with weights $\nu=\frac1{12},\frac14,\frac13,\frac14,\frac1{12}$. The NV centers inject polarization and, after being shelved off resonance, read out the local P1 polarization. Its diffusive tail gives $D$ for three groups, and for a $\nu=\frac14$ group while the other $\nu=\frac14$ group is driven. We compare with $gD$, where $g=2\pi^{1/3}$ corrects for the non-Gaussian polarization profile~\cite{zu2021}. We proceed in three steps and repeat them for the surface below.
	
	(i)~\textit{Non-Ising width.} The NV echo gives $\gamma=0.5~\mu$s$^{-1}$, and the $^{13}$C nuclei contribute $x_s\approx0.1$--0.3~MHz~\cite{zu2021}, so that $(\gamma+x_p)/2\pi=0.28$--0.68~MHz.
	
	(ii)~\textit{Ising width.} At the P1 density of Ref.~\cite{zu2021} ($110\pm10$~ppm),Equation~\eqref{eq:GammaL} gives $\Gamma_L/2\pi=2.55$~MHz without free parameters. The NV centers see the same P1 field, so their spectrum should be this Lorentzian, of full width 5.11~MHz. The measured spectrum~\cite{zu2021} is Lorentzian with a full width of 5.05~MHz [Fig.~\ref{fig:2}(a)]; a Gaussian of equal width leaves twice the residual, and fixing the Lorentzian width by Eq.~\eqref{eq:GammaL} fits as well as leaving it free. Hence $2\Gamma_L/(\gamma+x_p)=7.5$--18, and the P1 bath lies deep in the Ising blockade.
	
	(iii)~\textit{Removing the Ising term at fixed $J_{ij}$.} We solve the master equation exactly on networks of $\approx5000$ transport spins at the experimental density, with all five groups as Ising sources, rates~\eqref{eq:FGR} averaged over bath configurations, and $^{13}$C as static pair disorder~\cite{SM}. Fitting $\langle r^2\rangle$ over the experimental window of 30--300~$\mu$s gives $D=0.82\pm0.11$, $0.78\pm0.11$, $0.66\pm0.09$, $0.75\pm0.10$, and $0.19\pm0.04$~nm$^2/\mu$s for $\nu=\frac13$ (sample S1), $\frac13$ (S2, 25~K), $\frac14$, $\frac14$ driven, and $\frac1{12}$, compared with the measured $0.82\pm0.17$, $1.03\pm0.13$, $0.74\pm0.18$, $0.95\pm0.26$, and $0.33\pm0.08$~nm$^2/\mu$s [Fig.~\ref{fig:2}(b)]. Three of the five agree within $1\sigma$ and all within $2\sigma$, with no fitted parameter. Removing the Ising term at the same $J_{ij}$ raises $D$ to 4--11~nm$^2/\mu$s for the dense groups, five to fifteen times the measured values for $x_s=0.1$--0.3~MHz; matching the data without the Ising term would require $c\approx0.1$--0.3 instead of 1.94.

	With the Ising term, $D$ depends only weakly on the uncertain $^{13}$C width ($0.57\to0.52$~nm$^2/\mu$s for $x_s=0.1\to0.3$~MHz at $\nu=\frac14$) and on $\gamma$, as Eq.~\eqref{eq:Ddeep} requires, whereas the slowdown itself depends on $x_s$ (19, 7.7, and 2.7 for $x_s=0.1$, 0.3, and 1.0~MHz), so the absolute $D$ is the relevant test. Inverting Eq.~\eqref{eq:D} with the measured $D$ gives windows of 6.0, 4.6, and 2.4~MHz for $\nu=\frac13,\frac14,\frac1{12}$, compared with the predicted $(\gamma+x_p+2\Gamma_L)/2\pi=5.4$~MHz and with 0.28--0.68~MHz without the Ising term, so transport and spectroscopy measure the same Ising window. Driving the other $\nu=\frac14$ group at $\Omega/2\pi=11.7$~MHz removes only part of its field, since 28\% of it survives at this Rabi frequency~\cite{SM}; the network then predicts a speed-up of 1.16--1.20, consistent with the measured $1.32\pm0.48$. The two largest deviations, S2 ($1.9\sigma$) and the sparse $\nu=\frac1{12}$ group ($1.8\sigma$), have the same sign as the other residuals, and a $\nu$-independent channel of $0.14\pm0.06$~nm$^2/\mu$s, absent from both our network and the semiclassical model of Ref.~\cite{zu2021}, would account for all of them.
	
	\begin{figure*}[t]
		\centering
		\includegraphics[width=0.32\textwidth]{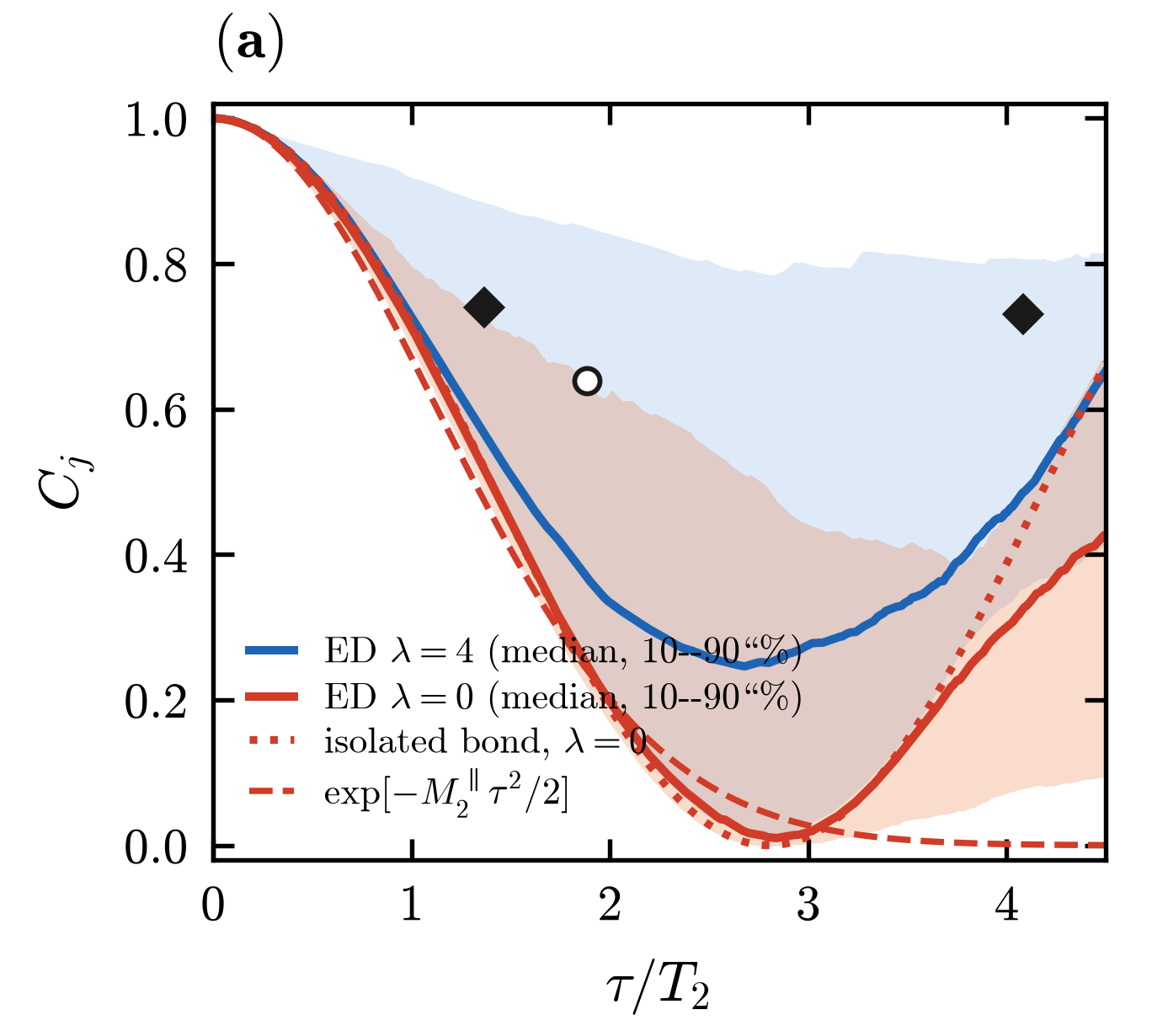}\hfill
		\includegraphics[width=0.32\textwidth]{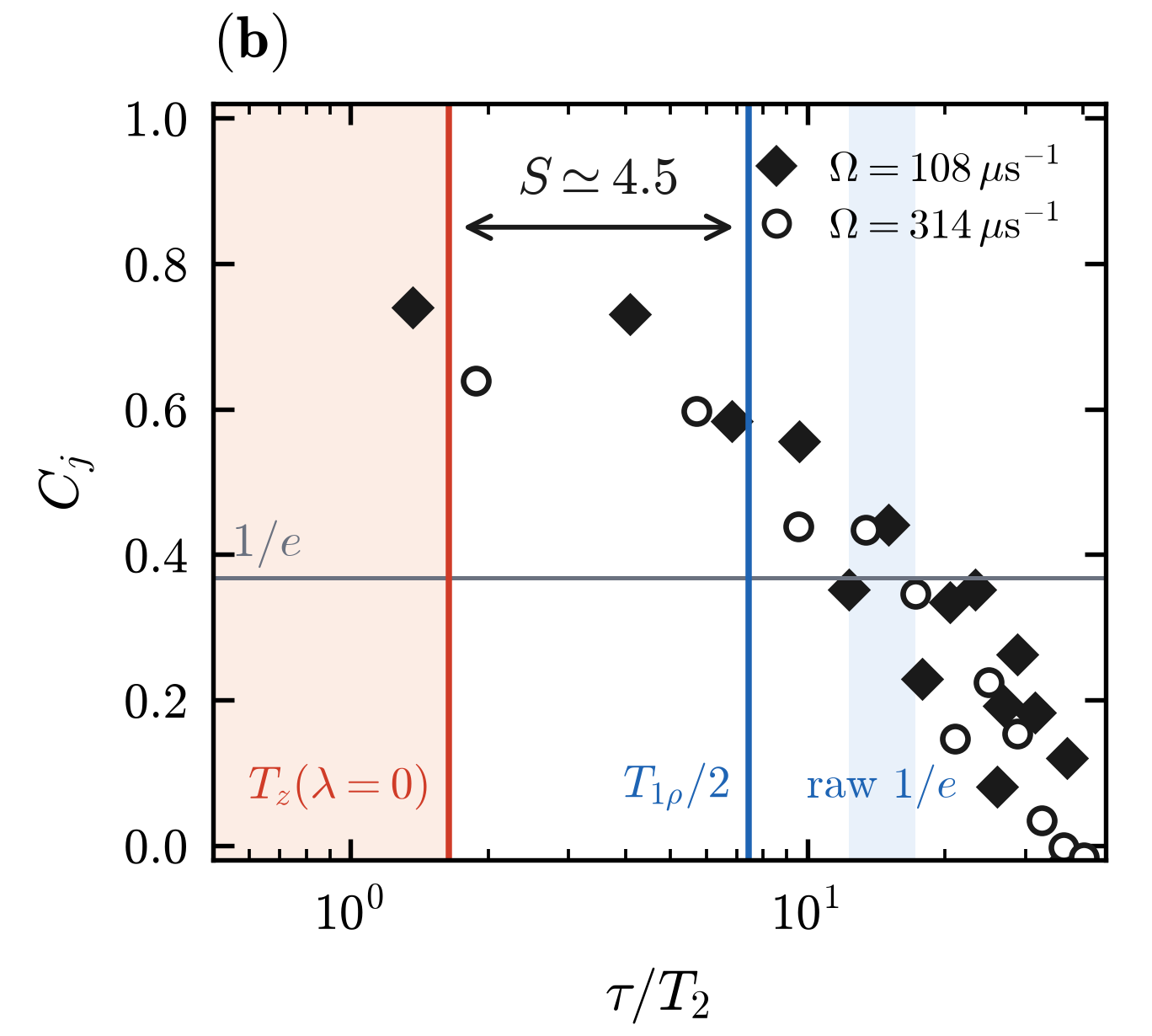}\hfill
		\includegraphics[width=0.32\textwidth]{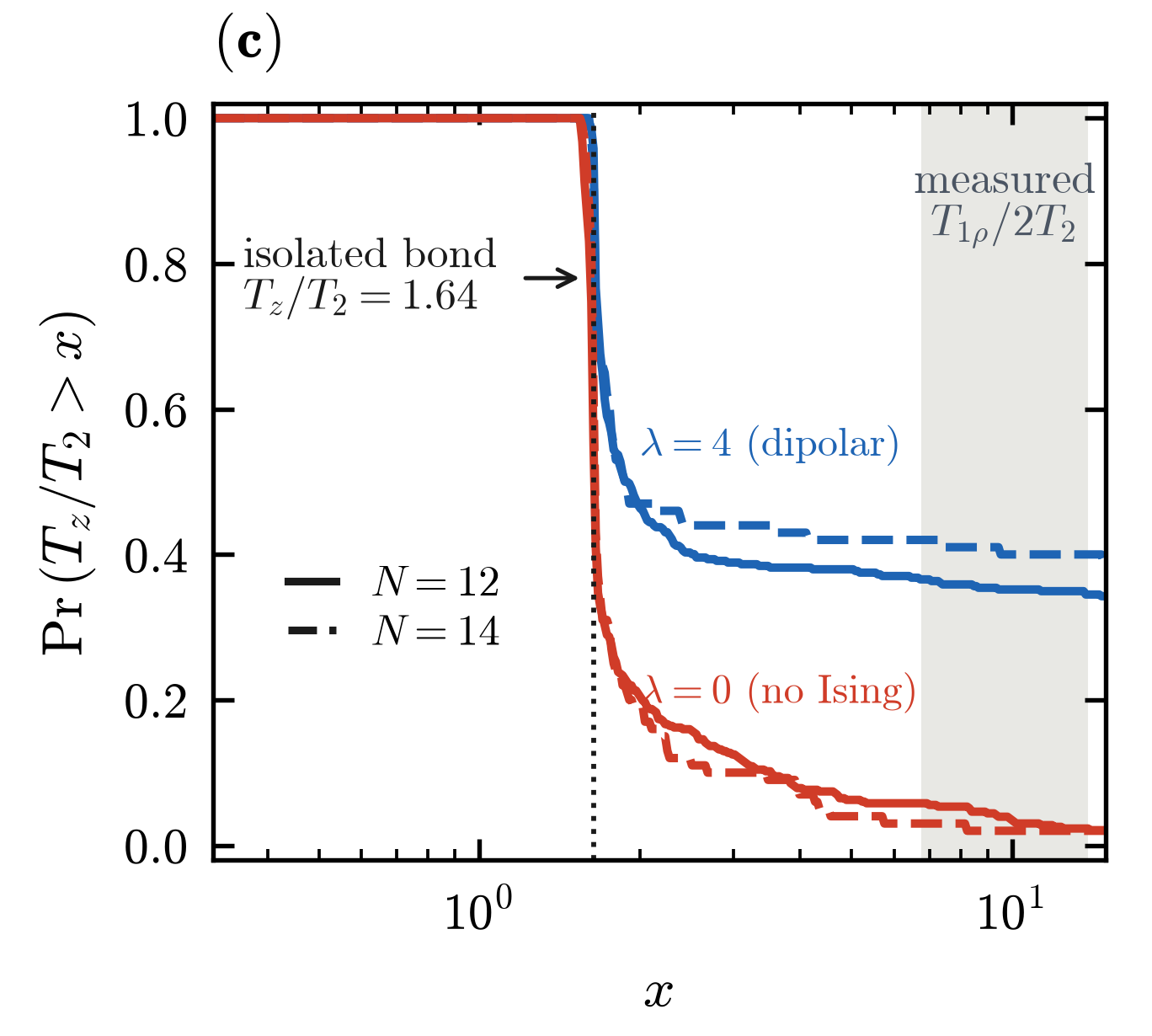}
		\caption{\textbf{The spin-lock plateau is an Ising blockade.}
			(a) Autocorrelation $C_j(\tau)$ of $S_j^z$ under $H_{\rm dd}$ from exact
			diagonalization of random two-dimensional clusters, with ($\lambda = 4$, blue) and
			without ($\lambda = 0$, red) the Ising term. Lines give the median and bands the
			10--90\% range. Time is in units of the Hahn-echo time $T_2$ of each dipolar
			configuration, which fixes $\sum_i J_{ij}^2$ through the moment identity; the
			$\lambda = 0$ curves use the same $J_{ij}$. Dotted: isolated bond at $\lambda = 0$.
			Dashed: Gaussian moment estimate $\exp(-M_2^\perp \tau^2/2)$. Symbols: measured decay
			of (b) at half the lock time.
			(b) Measured spin-lock decay of surface system~3 of Table~\ref{tab:inputs}~\cite{rezai2025} at
			$\Omega = 108$ (diamonds) and $314~\mu\mathrm{s}^{-1}$ (circles); the two drives give
			the same decay. Shaded band: raw $1/e$ crossing, $\tau = T_{1\rho}$. Vertical lines
			mark $T_z(\lambda=0) = 1.64\,T_2$ (red) and $T_{1\rho}/2$ (blue), whose ratio is
			$S \simeq 4.5$ [Eq.~\eqref{eq:S}].
			(c) Fraction of spins with $T_z/T_2 > x$ for $N = 12$ (solid) and $N = 14$ (dashed).
			Without the Ising term (red), the distribution collapses at the isolated-bond value
			$1.64$ (dotted line), and only 3--6\% of spins exceed $7T_2$. With it (blue),
			36--42\% do, and the tail saturates at the third of spins that do not reach $1/e$
			within the simulated time. Gray band: measured $T_{1\rho}/2T_2$ of the four systems.}
		\label{fig:3}
	\end{figure*}

	\textit{Test 2: the spin-lock plateau of a surface spin.}---Rezai \textit{et al.}~\cite{rezai2025} used shallow NV centers to probe single dominant electron spins on a (100) diamond surface with $\mathbf B\parallel[111]$, where the in-plane angular factor is $a=-\cos2\phi$. Under a spin lock of Rabi frequency $\Omega$, the central spin relaxes on a time $T_{1\rho}$ that is constant within $\pm20\%$ over $\Omega=79$--$419~\mu$s$^{-1}$ and long compared with the Hahn-echo time, with $T_{1\rho}/2T_2=6.8$--13.9 for four systems (Table~\ref{tab:inputs}). Their resonance-counting model, which already includes the Ising field as disorder, has no finite solution in this plateau~\cite{SM}. We repeat the three steps.
	
	(i)~\textit{Non-Ising width.} The drive removes it. In the toggling frame of the lock, an on-site field $\delta_j$ survives only as $\delta_j^2/2\Omega$, while $H_{\rm dd}$ keeps its secular form about the lock axis with all couplings halved, so $\lambda$ remains 4. A global rotation then maps the locked dynamics exactly onto the undriven dynamics at $\delta\equiv0$ with the time halved, $T_{1\rho}/2=T_z|_{\delta\equiv0}+O(\Omega^{-1})$ [End Matter~B]. All other channels (the proton field, static Larmor offsets, the NV field, and drive inhomogeneity) scale with a power of $\Omega$ and would change by at least a factor of 5.3 across the plateau, and lattice relaxation contributes at most 17--31\% [End Matter~B]. In the plateau, $\gamma+x_p\to0$.
	
	(ii)~\textit{Ising width.} At $\delta\equiv0$, $\eta_{ij}$ is the only detuning of a flip-flop, and it vanishes identically at $\lambda=0$ [Eq.~\eqref{eq:identity}]. It follows the two-dimensional law of Eq.~\eqref{eq:stable}, and Eq.~\eqref{eq:regime} gives 0.53 for the strongest bond of a typical spin, so the blockade competes directly with coherent exchange and the dynamics cannot be described by rates.
	
(iii)~\textit{Removing the Ising term at fixed $J_{ij}$.} We therefore compare exact dynamics. Cluster dynamical mean-field theory has captured such widely separated relaxation times~\cite{grasser2023}; our addition is the comparison at fixed geometry and couplings, which isolates the Ising term. The Hahn echo of Ref.~\cite{rezai2025} is free evolution under $H_{\rm dd}$, and the moment identity $M_2^\parallel/M_2^\perp=2/(1+\lambda^2/4)=2/5$ fixes $\sum_iJ_{ij}^2$ from the measured $T_2$ for each configuration~\cite{SM}. Without the Ising term, flip-flops alone would relax the spin in $T_z(\lambda{=}0)\simeq1.64\,T_2$. The exact isolated bond gives 1.639, the Gaussian moment estimate $\sqrt{5/2}=1.58$, and exact diagonalization of random two-dimensional clusters ($N=12$, 14) a median of 1.64, with only 3--6\% of the spins slower than $7\,T_2$ [Fig.~\ref{fig:3}(c)]. The measured plateau therefore gives
	\begin{equation}
		S\equiv\frac{T_z(\lambda=4)}{T_z(\lambda=0)}\bigg|_{\delta\equiv0}=\frac{T_{1\rho}/2T_2}{1.64}\simeq4\text{--}9
		\label{eq:S}
	\end{equation}
	for the four systems. Since the Ising term leaves the flip-flop second moment unchanged and acts only through $\eta_{ij}$, the whole excess $S-1$ is due to the Ising blockade. With the Ising term, the same clusters become bimodal: 36--42\% of the spins exceed $7\,T_2$, and one third do not reach $1/e$ within the simulated time [Fig.~\ref{fig:3}(c)]. At early times the measured decay lies at the 90th percentile of the $\lambda=0$ ensemble and well inside the $\lambda=4$ ensemble [Fig.~\ref{fig:3}(a)], but the relaxation time gives the clearer comparison [Fig.~\ref{fig:3}(b)].
	
	As expected from Eq.~\eqref{eq:regime}, the blockade affects only a subpopulation. The probability that the strongest bond of a spin cannot bring its autocorrelation below $1/e$ is $p_B=0.47$ for randomly placed spins, and $p_B=0.58$--0.73 when, as in Ref.~\cite{rezai2025}, systems with a single dominant surface spin are selected~\cite{SM}. Equation~\eqref{eq:S} therefore refers to individual spins.
	
	\textit{Comparison of platforms.}---Figure~\ref{fig:crossover} places the platforms on the scale of Eq.~\eqref{eq:criterion}. In dense NV ensembles the NV-NV Ising field is weak compared with the static single-spin line. At the density of the superradiant maser of Ref.~\cite{kersten2026}, $2\Gamma_L/2\pi\approx0.46$~MHz against $x_s/2\pi=4.3$~MHz, and the blockade lengthens spectral-hole refilling by only 3--4\%~\cite{SM}, so the tenfold discrepancy reported there must have a different origin. The P1 bath lies an order of magnitude on the other side. On a locked surface the drive removes $x_p$ and $\gamma$ completely, and the blockade is limited only by coherent exchange [Eq.~\eqref{eq:regime}].

	Global Hamiltonian engineering cannot remove the blockade. For like spin-$\frac12$ dipoles, any global control sequence leaves the averaged coupling tensor symmetric and traceless, and U(1) symmetry about an axis then forces $Q=-4J$, i.e., $\lambda=4$, or zero coupling [End Matter~C; cf.\ Ref.~\cite{choi2020prx}]. The spin lock is one example. The $\lambda=0$ dynamics is thus only a theoretical reference, and the Ising blockade is intrinsic to magnetic dipolar transport in such systems.
	
	\textit{Predictions.}---The theory makes the following parameter-free predictions. (i)~At fixed $\nu$, $D\propto\rho^{1/3}$ above $\rho^*\approx6$--15~ppm and $D\propto\rho^{4/3}$ below, while the NV line stays Lorentzian with a full width of 0.046~MHz per ppm of P1 [Eq.~\eqref{eq:GammaL}]. (ii)~Deep in the blockade, the product of $D$ and the NV linewidth stays constant when Ising sources are removed by driving. (iii)~Driving the three P1 groups lying at least 85~MHz away from a $\nu=\frac14$ transport group removes two thirds of its Ising sources and speeds up diffusion by a factor of 1.8 at $\Omega/2\pi=11.7$~MHz and 2.7 under strong drive. (iv)~$D$ is unchanged in $^{12}$C-purified diamond and decreases only when $^{13}$C enrichment raises $x_p$ to $2\Gamma_L$. (v)~At $\delta\equiv0$ all surface time scales are proportional to $1/J_0\rho^{3/2}$, so the distribution of $T_{1\rho}/2T_2$ is the same for any surface density, with about half of randomly placed spins blockaded.
	
	\textit{Conclusion.}---The Ising interaction moves no polarization, yet it sets how fast polarization moves. It acts exactly through the pair detuning [Eq.~\eqref{eq:identity}], whose universal statistics are generated by the same couplings that drive transport [Eqs.~\eqref{eq:stable}--\eqref{eq:regime}], so dipolar transport limits itself. Removing the Ising term at fixed flip-flop couplings accounts, without fitting, for spin diffusion in the P1 bath and for the relaxation plateau of a locked surface spin, and Eq.~\eqref{eq:criterion} identifies where the blockade is negligible. Because global control cannot tune $\lambda$ in a spin-$\frac12$ species, the Ising blockade is an intrinsic limit on polarization transport, spectral-hole refilling, and thermalization in dipolar quantum platforms; selective driving of Ising sources~\cite{zu2021} can weaken it.
	\bibliography{prl,sm}

\begin{thebibliography}{40}%
\makeatletter
\providecommand \@ifxundefined [1]{%
 \@ifx{#1\undefined}
}%
\providecommand \@ifnum [1]{%
 \ifnum #1\expandafter \@firstoftwo
 \else \expandafter \@secondoftwo
 \fi
}%
\providecommand \@ifx [1]{%
 \ifx #1\expandafter \@firstoftwo
 \else \expandafter \@secondoftwo
 \fi
}%
\providecommand \natexlab [1]{#1}%
\providecommand \enquote  [1]{``#1''}%
\providecommand \bibnamefont  [1]{#1}%
\providecommand \bibfnamefont [1]{#1}%
\providecommand \citenamefont [1]{#1}%
\providecommand \href@noop [0]{\@secondoftwo}%
\providecommand \href [0]{\begingroup \@sanitize@url \@href}%
\providecommand \@href[1]{\@@startlink{#1}\@@href}%
\providecommand \@@href[1]{\endgroup#1\@@endlink}%
\providecommand \@sanitize@url [0]{\catcode `\\12\catcode `\$12\catcode
  `\&12\catcode `\#12\catcode `\^12\catcode `\_12\catcode `\%12\relax}%
\providecommand \@@startlink[1]{}%
\providecommand \@@endlink[0]{}%
\providecommand \url  [0]{\begingroup\@sanitize@url \@url }%
\providecommand \@url [1]{\endgroup\@href {#1}{\urlprefix }}%
\providecommand \urlprefix  [0]{URL }%
\providecommand \Eprint [0]{\href }%
\providecommand \doibase [0]{https://doi.org/}%
\providecommand \selectlanguage [0]{\@gobble}%
\providecommand \bibinfo  [0]{\@secondoftwo}%
\providecommand \bibfield  [0]{\@secondoftwo}%
\providecommand \translation [1]{[#1]}%
\providecommand \BibitemOpen [0]{}%
\providecommand \bibitemStop [0]{}%
\providecommand \bibitemNoStop [0]{.\EOS\space}%
\providecommand \EOS [0]{\spacefactor3000\relax}%
\providecommand \BibitemShut  [1]{\csname bibitem#1\endcsname}%
\let\auto@bib@innerbib\@empty
\bibitem [{\citenamefont {Degen}\ \emph {et~al.}(2017)\citenamefont {Degen},
  \citenamefont {Reinhard},\ and\ \citenamefont {Cappellaro}}]{degen2017}%
  \BibitemOpen
  \bibfield  {author} {\bibinfo {author} {\bibfnamefont {C.~L.}\ \bibnamefont
  {Degen}}, \bibinfo {author} {\bibfnamefont {F.}~\bibnamefont {Reinhard}},\
  and\ \bibinfo {author} {\bibfnamefont {P.}~\bibnamefont {Cappellaro}},\
  }\bibfield  {title} {\bibinfo {title} {Quantum sensing},\ }\href
  {https://doi.org/10.1103/RevModPhys.89.035002} {\bibfield  {journal}
  {\bibinfo  {journal} {Reviews of Modern Physics}\ }\textbf {\bibinfo {volume}
  {89}},\ \bibinfo {pages} {035002} (\bibinfo {year} {2017})}\BibitemShut
  {NoStop}%
\bibitem [{\citenamefont {Zhou}\ \emph {et~al.}(2020)\citenamefont {Zhou},
  \citenamefont {Choi}, \citenamefont {Choi}, \citenamefont {Landig},
  \citenamefont {Douglas}, \citenamefont {Isoya}, \citenamefont {Jelezko},
  \citenamefont {Onoda}, \citenamefont {Sumiya}, \citenamefont {Cappellaro},
  \citenamefont {Knowles}, \citenamefont {Park},\ and\ \citenamefont
  {Lukin}}]{zhou2020}%
  \BibitemOpen
  \bibfield  {author} {\bibinfo {author} {\bibfnamefont {H.}~\bibnamefont
  {Zhou}}, \bibinfo {author} {\bibfnamefont {J.}~\bibnamefont {Choi}}, \bibinfo
  {author} {\bibfnamefont {S.}~\bibnamefont {Choi}}, \bibinfo {author}
  {\bibfnamefont {R.}~\bibnamefont {Landig}}, \bibinfo {author} {\bibfnamefont
  {A.~M.}\ \bibnamefont {Douglas}}, \bibinfo {author} {\bibfnamefont
  {J.}~\bibnamefont {Isoya}}, \bibinfo {author} {\bibfnamefont
  {F.}~\bibnamefont {Jelezko}}, \bibinfo {author} {\bibfnamefont
  {S.}~\bibnamefont {Onoda}}, \bibinfo {author} {\bibfnamefont
  {H.}~\bibnamefont {Sumiya}}, \bibinfo {author} {\bibfnamefont
  {P.}~\bibnamefont {Cappellaro}}, \bibinfo {author} {\bibfnamefont {H.~S.}\
  \bibnamefont {Knowles}}, \bibinfo {author} {\bibfnamefont {H.}~\bibnamefont
  {Park}},\ and\ \bibinfo {author} {\bibfnamefont {M.~D.}\ \bibnamefont
  {Lukin}},\ }\bibfield  {title} {\bibinfo {title} {Quantum {{Metrology}} with
  {{Strongly Interacting Spin Systems}}},\ }\href
  {https://doi.org/10.1103/PhysRevX.10.031003} {\bibfield  {journal} {\bibinfo
  {journal} {Physical Review X}\ }\textbf {\bibinfo {volume} {10}},\ \bibinfo
  {pages} {031003} (\bibinfo {year} {2020})}\BibitemShut {NoStop}%
\bibitem [{\citenamefont {Wu}\ \emph {et~al.}(2025)\citenamefont {Wu},
  \citenamefont {Davis}, \citenamefont {Hughes}, \citenamefont {Ye},
  \citenamefont {Wang}, \citenamefont {Kufel}, \citenamefont {Ono},
  \citenamefont {Meynell}, \citenamefont {Block}, \citenamefont {Liu},
  \citenamefont {Yang}, \citenamefont {Bleszynski~Jayich},\ and\ \citenamefont
  {Yao}}]{wu2025}%
  \BibitemOpen
  \bibfield  {author} {\bibinfo {author} {\bibfnamefont {W.}~\bibnamefont
  {Wu}}, \bibinfo {author} {\bibfnamefont {E.~J.}\ \bibnamefont {Davis}},
  \bibinfo {author} {\bibfnamefont {L.~B.}\ \bibnamefont {Hughes}}, \bibinfo
  {author} {\bibfnamefont {B.}~\bibnamefont {Ye}}, \bibinfo {author}
  {\bibfnamefont {Z.}~\bibnamefont {Wang}}, \bibinfo {author} {\bibfnamefont
  {D.}~\bibnamefont {Kufel}}, \bibinfo {author} {\bibfnamefont
  {T.}~\bibnamefont {Ono}}, \bibinfo {author} {\bibfnamefont {S.~A.}\
  \bibnamefont {Meynell}}, \bibinfo {author} {\bibfnamefont {M.}~\bibnamefont
  {Block}}, \bibinfo {author} {\bibfnamefont {C.}~\bibnamefont {Liu}}, \bibinfo
  {author} {\bibfnamefont {H.}~\bibnamefont {Yang}}, \bibinfo {author}
  {\bibfnamefont {A.~C.}\ \bibnamefont {Bleszynski~Jayich}},\ and\ \bibinfo
  {author} {\bibfnamefont {N.~Y.}\ \bibnamefont {Yao}},\ }\bibfield  {title}
  {\bibinfo {title} {Spin squeezing in an ensemble of nitrogen--vacancy centres
  in diamond},\ }\href {https://doi.org/10.1038/s41586-025-09524-8} {\bibfield
  {journal} {\bibinfo  {journal} {Nature}\ }\textbf {\bibinfo {volume} {646}},\
  \bibinfo {pages} {74} (\bibinfo {year} {2025})}\BibitemShut {NoStop}%
\bibitem [{\citenamefont {Wolfowicz}\ \emph {et~al.}(2021)\citenamefont
  {Wolfowicz}, \citenamefont {Heremans}, \citenamefont {Anderson},
  \citenamefont {Kanai}, \citenamefont {Seo}, \citenamefont {Gali},
  \citenamefont {Galli},\ and\ \citenamefont {Awschalom}}]{wolfowicz2021}%
  \BibitemOpen
  \bibfield  {author} {\bibinfo {author} {\bibfnamefont {G.}~\bibnamefont
  {Wolfowicz}}, \bibinfo {author} {\bibfnamefont {F.~J.}\ \bibnamefont
  {Heremans}}, \bibinfo {author} {\bibfnamefont {C.~P.}\ \bibnamefont
  {Anderson}}, \bibinfo {author} {\bibfnamefont {S.}~\bibnamefont {Kanai}},
  \bibinfo {author} {\bibfnamefont {H.}~\bibnamefont {Seo}}, \bibinfo {author}
  {\bibfnamefont {A.}~\bibnamefont {Gali}}, \bibinfo {author} {\bibfnamefont
  {G.}~\bibnamefont {Galli}},\ and\ \bibinfo {author} {\bibfnamefont {D.~D.}\
  \bibnamefont {Awschalom}},\ }\bibfield  {title} {\bibinfo {title} {Quantum
  guidelines for solid-state spin defects},\ }\href
  {https://doi.org/10.1038/s41578-021-00306-y} {\bibfield  {journal} {\bibinfo
  {journal} {Nature Reviews Materials}\ }\textbf {\bibinfo {volume} {6}},\
  \bibinfo {pages} {906} (\bibinfo {year} {2021})}\BibitemShut {NoStop}%
\bibitem [{\citenamefont {Hall}\ \emph {et~al.}(2016)\citenamefont {Hall},
  \citenamefont {Kehayias}, \citenamefont {Simpson}, \citenamefont {Jarmola},
  \citenamefont {Stacey}, \citenamefont {Budker},\ and\ \citenamefont
  {Hollenberg}}]{hallDetectionNanoscaleElectron2016}%
  \BibitemOpen
  \bibfield  {author} {\bibinfo {author} {\bibfnamefont {L.~T.}\ \bibnamefont
  {Hall}}, \bibinfo {author} {\bibfnamefont {P.}~\bibnamefont {Kehayias}},
  \bibinfo {author} {\bibfnamefont {D.~A.}\ \bibnamefont {Simpson}}, \bibinfo
  {author} {\bibfnamefont {A.}~\bibnamefont {Jarmola}}, \bibinfo {author}
  {\bibfnamefont {A.}~\bibnamefont {Stacey}}, \bibinfo {author} {\bibfnamefont
  {D.}~\bibnamefont {Budker}},\ and\ \bibinfo {author} {\bibfnamefont
  {L.~C.~L.}\ \bibnamefont {Hollenberg}},\ }\bibfield  {title} {\bibinfo
  {title} {Detection of nanoscale electron spin resonance spectra demonstrated
  using nitrogen-vacancy centre probes in diamond},\ }\href
  {https://doi.org/10.1038/ncomms10211} {\bibfield  {journal} {\bibinfo
  {journal} {Nature Communications}\ }\textbf {\bibinfo {volume} {7}},\
  \bibinfo {pages} {10211} (\bibinfo {year} {2016})}\BibitemShut {NoStop}%
\bibitem [{\citenamefont {Putz}\ \emph {et~al.}(2017)\citenamefont {Putz},
  \citenamefont {Angerer}, \citenamefont {Krimer}, \citenamefont {Glattauer},
  \citenamefont {Munro}, \citenamefont {Rotter}, \citenamefont {Schmiedmayer},\
  and\ \citenamefont {Majer}}]{putzSpectralHoleBurning2017}%
  \BibitemOpen
  \bibfield  {author} {\bibinfo {author} {\bibfnamefont {S.}~\bibnamefont
  {Putz}}, \bibinfo {author} {\bibfnamefont {A.}~\bibnamefont {Angerer}},
  \bibinfo {author} {\bibfnamefont {D.~O.}\ \bibnamefont {Krimer}}, \bibinfo
  {author} {\bibfnamefont {R.}~\bibnamefont {Glattauer}}, \bibinfo {author}
  {\bibfnamefont {W.~J.}\ \bibnamefont {Munro}}, \bibinfo {author}
  {\bibfnamefont {S.}~\bibnamefont {Rotter}}, \bibinfo {author} {\bibfnamefont
  {J.}~\bibnamefont {Schmiedmayer}},\ and\ \bibinfo {author} {\bibfnamefont
  {J.}~\bibnamefont {Majer}},\ }\bibfield  {title} {\bibinfo {title} {Spectral
  hole burning and its application in microwave photonics},\ }\href
  {https://doi.org/10.1038/nphoton.2016.225} {\bibfield  {journal} {\bibinfo
  {journal} {Nature Photonics}\ }\textbf {\bibinfo {volume} {11}},\ \bibinfo
  {pages} {36} (\bibinfo {year} {2017})}\BibitemShut {NoStop}%
\bibitem [{\citenamefont {Angerer}\ \emph {et~al.}(2018)\citenamefont
  {Angerer}, \citenamefont {Streltsov}, \citenamefont {Astner}, \citenamefont
  {Putz}, \citenamefont {Sumiya}, \citenamefont {Onoda}, \citenamefont {Isoya},
  \citenamefont {Munro}, \citenamefont {Nemoto}, \citenamefont {Schmiedmayer},\
  and\ \citenamefont {Majer}}]{angerer2018}%
  \BibitemOpen
  \bibfield  {author} {\bibinfo {author} {\bibfnamefont {A.}~\bibnamefont
  {Angerer}}, \bibinfo {author} {\bibfnamefont {K.}~\bibnamefont {Streltsov}},
  \bibinfo {author} {\bibfnamefont {T.}~\bibnamefont {Astner}}, \bibinfo
  {author} {\bibfnamefont {S.}~\bibnamefont {Putz}}, \bibinfo {author}
  {\bibfnamefont {H.}~\bibnamefont {Sumiya}}, \bibinfo {author} {\bibfnamefont
  {S.}~\bibnamefont {Onoda}}, \bibinfo {author} {\bibfnamefont
  {J.}~\bibnamefont {Isoya}}, \bibinfo {author} {\bibfnamefont {W.~J.}\
  \bibnamefont {Munro}}, \bibinfo {author} {\bibfnamefont {K.}~\bibnamefont
  {Nemoto}}, \bibinfo {author} {\bibfnamefont {J.}~\bibnamefont
  {Schmiedmayer}},\ and\ \bibinfo {author} {\bibfnamefont {J.}~\bibnamefont
  {Majer}},\ }\bibfield  {title} {\bibinfo {title} {Superradiant emission from
  colour centres in diamond},\ }\href
  {https://doi.org/10.1038/s41567-018-0269-7} {\bibfield  {journal} {\bibinfo
  {journal} {Nature Physics}\ }\textbf {\bibinfo {volume} {14}},\ \bibinfo
  {pages} {1168} (\bibinfo {year} {2018})}\BibitemShut {NoStop}%
\bibitem [{\citenamefont {Breeze}\ \emph {et~al.}(2018)\citenamefont {Breeze},
  \citenamefont {Salvadori}, \citenamefont {Sathian}, \citenamefont {Alford},\
  and\ \citenamefont {Kay}}]{breezeContinuouswaveRoomtemperatureDiamond2018}%
  \BibitemOpen
  \bibfield  {author} {\bibinfo {author} {\bibfnamefont {J.~D.}\ \bibnamefont
  {Breeze}}, \bibinfo {author} {\bibfnamefont {E.}~\bibnamefont {Salvadori}},
  \bibinfo {author} {\bibfnamefont {J.}~\bibnamefont {Sathian}}, \bibinfo
  {author} {\bibfnamefont {N.~M.}\ \bibnamefont {Alford}},\ and\ \bibinfo
  {author} {\bibfnamefont {C.~W.~M.}\ \bibnamefont {Kay}},\ }\bibfield  {title}
  {\bibinfo {title} {Continuous-wave room-temperature diamond maser},\ }\href
  {https://doi.org/10.1038/nature25970} {\bibfield  {journal} {\bibinfo
  {journal} {Nature}\ }\textbf {\bibinfo {volume} {555}},\ \bibinfo {pages}
  {493} (\bibinfo {year} {2018})}\BibitemShut {NoStop}%
\bibitem [{\citenamefont {Kersten}\ \emph {et~al.}(2023)\citenamefont
  {Kersten}, \citenamefont {De~Zordo}, \citenamefont {Diekmann}, \citenamefont
  {Reiter}, \citenamefont {Zens}, \citenamefont {Kanagin}, \citenamefont
  {Rotter}, \citenamefont {Schmiedmayer},\ and\ \citenamefont
  {Angerer}}]{kersten2023}%
  \BibitemOpen
  \bibfield  {author} {\bibinfo {author} {\bibfnamefont {W.}~\bibnamefont
  {Kersten}}, \bibinfo {author} {\bibfnamefont {N.}~\bibnamefont {De~Zordo}},
  \bibinfo {author} {\bibfnamefont {O.}~\bibnamefont {Diekmann}}, \bibinfo
  {author} {\bibfnamefont {T.}~\bibnamefont {Reiter}}, \bibinfo {author}
  {\bibfnamefont {M.}~\bibnamefont {Zens}}, \bibinfo {author} {\bibfnamefont
  {A.~N.}\ \bibnamefont {Kanagin}}, \bibinfo {author} {\bibfnamefont
  {S.}~\bibnamefont {Rotter}}, \bibinfo {author} {\bibfnamefont
  {J.}~\bibnamefont {Schmiedmayer}},\ and\ \bibinfo {author} {\bibfnamefont
  {A.}~\bibnamefont {Angerer}},\ }\bibfield  {title} {\bibinfo {title}
  {Triggered {{Superradiance}} and {{Spin Inversion Storage}} in a {{Hybrid
  Quantum System}}},\ }\href {https://doi.org/10.1103/PhysRevLett.131.043601}
  {\bibfield  {journal} {\bibinfo  {journal} {Physical Review Letters}\
  }\textbf {\bibinfo {volume} {131}},\ \bibinfo {pages} {043601} (\bibinfo
  {year} {2023})}\BibitemShut {NoStop}%
\bibitem [{\citenamefont {Kersten}\ \emph {et~al.}(2026)\citenamefont
  {Kersten}, \citenamefont {De~Zordo}, \citenamefont {Diekmann}, \citenamefont
  {Redchenko}, \citenamefont {Kanagin}, \citenamefont {Angerer}, \citenamefont
  {Munro}, \citenamefont {Nemoto}, \citenamefont {Mazets}, \citenamefont
  {Rotter}, \citenamefont {Pohl},\ and\ \citenamefont
  {Schmiedmayer}}]{kersten2026}%
  \BibitemOpen
  \bibfield  {author} {\bibinfo {author} {\bibfnamefont {W.}~\bibnamefont
  {Kersten}}, \bibinfo {author} {\bibfnamefont {N.}~\bibnamefont {De~Zordo}},
  \bibinfo {author} {\bibfnamefont {O.}~\bibnamefont {Diekmann}}, \bibinfo
  {author} {\bibfnamefont {E.~S.}\ \bibnamefont {Redchenko}}, \bibinfo {author}
  {\bibfnamefont {A.~N.}\ \bibnamefont {Kanagin}}, \bibinfo {author}
  {\bibfnamefont {A.}~\bibnamefont {Angerer}}, \bibinfo {author} {\bibfnamefont
  {W.~J.}\ \bibnamefont {Munro}}, \bibinfo {author} {\bibfnamefont
  {K.}~\bibnamefont {Nemoto}}, \bibinfo {author} {\bibfnamefont {I.~E.}\
  \bibnamefont {Mazets}}, \bibinfo {author} {\bibfnamefont {S.}~\bibnamefont
  {Rotter}}, \bibinfo {author} {\bibfnamefont {T.}~\bibnamefont {Pohl}},\ and\
  \bibinfo {author} {\bibfnamefont {J.}~\bibnamefont {Schmiedmayer}},\
  }\bibfield  {title} {\bibinfo {title} {Self-induced superradiant masing},\
  }\href {https://doi.org/10.1038/s41567-025-03123-0} {\bibfield  {journal}
  {\bibinfo  {journal} {Nature Physics}\ }\textbf {\bibinfo {volume} {22}},\
  \bibinfo {pages} {158} (\bibinfo {year} {2026})}\BibitemShut {NoStop}%
\bibitem [{\citenamefont {Choi}\ \emph {et~al.}(2017)\citenamefont {Choi},
  \citenamefont {Choi}, \citenamefont {Kucsko}, \citenamefont {Maurer},
  \citenamefont {Shields}, \citenamefont {Sumiya}, \citenamefont {Onoda},
  \citenamefont {Isoya}, \citenamefont {Demler}, \citenamefont {Jelezko},
  \citenamefont {Yao},\ and\ \citenamefont {Lukin}}]{choi2017}%
  \BibitemOpen
  \bibfield  {author} {\bibinfo {author} {\bibfnamefont {J.}~\bibnamefont
  {Choi}}, \bibinfo {author} {\bibfnamefont {S.}~\bibnamefont {Choi}}, \bibinfo
  {author} {\bibfnamefont {G.}~\bibnamefont {Kucsko}}, \bibinfo {author}
  {\bibfnamefont {P.~C.}\ \bibnamefont {Maurer}}, \bibinfo {author}
  {\bibfnamefont {B.~J.}\ \bibnamefont {Shields}}, \bibinfo {author}
  {\bibfnamefont {H.}~\bibnamefont {Sumiya}}, \bibinfo {author} {\bibfnamefont
  {S.}~\bibnamefont {Onoda}}, \bibinfo {author} {\bibfnamefont
  {J.}~\bibnamefont {Isoya}}, \bibinfo {author} {\bibfnamefont
  {E.}~\bibnamefont {Demler}}, \bibinfo {author} {\bibfnamefont
  {F.}~\bibnamefont {Jelezko}}, \bibinfo {author} {\bibfnamefont {N.~Y.}\
  \bibnamefont {Yao}},\ and\ \bibinfo {author} {\bibfnamefont {M.~D.}\
  \bibnamefont {Lukin}},\ }\bibfield  {title} {\bibinfo {title} {Depolarization
  dynamics in a strongly interacting solid-state spin ensemble},\ }\href
  {https://doi.org/10.1103/PhysRevLett.118.093601} {\bibfield  {journal}
  {\bibinfo  {journal} {Physical Review Letters}\ }\textbf {\bibinfo {volume}
  {118}},\ \bibinfo {pages} {93601} (\bibinfo {year} {2017})}\BibitemShut
  {NoStop}%
\bibitem [{\citenamefont {Kucsko}\ \emph {et~al.}(2018)\citenamefont {Kucsko},
  \citenamefont {Choi}, \citenamefont {Choi}, \citenamefont {Maurer},
  \citenamefont {Zhou}, \citenamefont {Landig}, \citenamefont {Sumiya},
  \citenamefont {Onoda}, \citenamefont {Isoya}, \citenamefont {Jelezko},
  \citenamefont {Demler}, \citenamefont {Yao},\ and\ \citenamefont
  {Lukin}}]{kucsko2018}%
  \BibitemOpen
  \bibfield  {author} {\bibinfo {author} {\bibfnamefont {G.}~\bibnamefont
  {Kucsko}}, \bibinfo {author} {\bibfnamefont {S.}~\bibnamefont {Choi}},
  \bibinfo {author} {\bibfnamefont {J.}~\bibnamefont {Choi}}, \bibinfo {author}
  {\bibfnamefont {P.~C.}\ \bibnamefont {Maurer}}, \bibinfo {author}
  {\bibfnamefont {H.}~\bibnamefont {Zhou}}, \bibinfo {author} {\bibfnamefont
  {R.}~\bibnamefont {Landig}}, \bibinfo {author} {\bibfnamefont
  {H.}~\bibnamefont {Sumiya}}, \bibinfo {author} {\bibfnamefont
  {S.}~\bibnamefont {Onoda}}, \bibinfo {author} {\bibfnamefont
  {J.}~\bibnamefont {Isoya}}, \bibinfo {author} {\bibfnamefont
  {F.}~\bibnamefont {Jelezko}}, \bibinfo {author} {\bibfnamefont
  {E.}~\bibnamefont {Demler}}, \bibinfo {author} {\bibfnamefont {N.~Y.}\
  \bibnamefont {Yao}},\ and\ \bibinfo {author} {\bibfnamefont {M.~D.}\
  \bibnamefont {Lukin}},\ }\bibfield  {title} {\bibinfo {title} {Critical
  {{Thermalization}} of a {{Disordered Dipolar Spin System}} in {{Diamond}}},\
  }\href {https://doi.org/10.1103/PhysRevLett.121.023601} {\bibfield  {journal}
  {\bibinfo  {journal} {Physical Review Letters}\ }\textbf {\bibinfo {volume}
  {121}},\ \bibinfo {pages} {023601} (\bibinfo {year} {2018})}\BibitemShut
  {NoStop}%
\bibitem [{\citenamefont {Choi}\ \emph {et~al.}(2019)\citenamefont {Choi},
  \citenamefont {Zhou}, \citenamefont {Choi}, \citenamefont {Landig},
  \citenamefont {Ho}, \citenamefont {Isoya}, \citenamefont {Jelezko},
  \citenamefont {Onoda}, \citenamefont {Sumiya}, \citenamefont {Abanin},\ and\
  \citenamefont {Lukin}}]{choi2019}%
  \BibitemOpen
  \bibfield  {author} {\bibinfo {author} {\bibfnamefont {J.}~\bibnamefont
  {Choi}}, \bibinfo {author} {\bibfnamefont {H.}~\bibnamefont {Zhou}}, \bibinfo
  {author} {\bibfnamefont {S.}~\bibnamefont {Choi}}, \bibinfo {author}
  {\bibfnamefont {R.}~\bibnamefont {Landig}}, \bibinfo {author} {\bibfnamefont
  {W.~W.}\ \bibnamefont {Ho}}, \bibinfo {author} {\bibfnamefont
  {J.}~\bibnamefont {Isoya}}, \bibinfo {author} {\bibfnamefont
  {F.}~\bibnamefont {Jelezko}}, \bibinfo {author} {\bibfnamefont
  {S.}~\bibnamefont {Onoda}}, \bibinfo {author} {\bibfnamefont
  {H.}~\bibnamefont {Sumiya}}, \bibinfo {author} {\bibfnamefont {D.~A.}\
  \bibnamefont {Abanin}},\ and\ \bibinfo {author} {\bibfnamefont {M.~D.}\
  \bibnamefont {Lukin}},\ }\bibfield  {title} {\bibinfo {title} {Probing
  {{Quantum Thermalization}} of a {{Disordered Dipolar Spin Ensemble}} with
  {{Discrete Time-Crystalline Order}}},\ }\href
  {https://doi.org/10.1103/PhysRevLett.122.043603} {\bibfield  {journal}
  {\bibinfo  {journal} {Physical Review Letters}\ }\textbf {\bibinfo {volume}
  {122}},\ \bibinfo {pages} {043603} (\bibinfo {year} {2019})}\BibitemShut
  {NoStop}%
\bibitem [{\citenamefont {Zu}\ \emph {et~al.}(2021)\citenamefont {Zu},
  \citenamefont {Machado}, \citenamefont {Ye}, \citenamefont {Choi},
  \citenamefont {Kobrin}, \citenamefont {Mittiga}, \citenamefont {Hsieh},
  \citenamefont {Bhattacharyya}, \citenamefont {Markham}, \citenamefont
  {Twitchen}, \citenamefont {Jarmola}, \citenamefont {Budker}, \citenamefont
  {Laumann}, \citenamefont {Moore},\ and\ \citenamefont {Yao}}]{zu2021}%
  \BibitemOpen
  \bibfield  {author} {\bibinfo {author} {\bibfnamefont {C.}~\bibnamefont
  {Zu}}, \bibinfo {author} {\bibfnamefont {F.}~\bibnamefont {Machado}},
  \bibinfo {author} {\bibfnamefont {B.}~\bibnamefont {Ye}}, \bibinfo {author}
  {\bibfnamefont {S.}~\bibnamefont {Choi}}, \bibinfo {author} {\bibfnamefont
  {B.}~\bibnamefont {Kobrin}}, \bibinfo {author} {\bibfnamefont
  {T.}~\bibnamefont {Mittiga}}, \bibinfo {author} {\bibfnamefont
  {S.}~\bibnamefont {Hsieh}}, \bibinfo {author} {\bibfnamefont
  {P.}~\bibnamefont {Bhattacharyya}}, \bibinfo {author} {\bibfnamefont
  {M.}~\bibnamefont {Markham}}, \bibinfo {author} {\bibfnamefont
  {D.}~\bibnamefont {Twitchen}}, \bibinfo {author} {\bibfnamefont
  {A.}~\bibnamefont {Jarmola}}, \bibinfo {author} {\bibfnamefont
  {D.}~\bibnamefont {Budker}}, \bibinfo {author} {\bibfnamefont {C.~R.}\
  \bibnamefont {Laumann}}, \bibinfo {author} {\bibfnamefont {J.~E.}\
  \bibnamefont {Moore}},\ and\ \bibinfo {author} {\bibfnamefont {N.~Y.}\
  \bibnamefont {Yao}},\ }\bibfield  {title} {\bibinfo {title} {Emergent
  hydrodynamics in a strongly interacting dipolar spin ensemble},\ }\href
  {https://doi.org/10.1038/s41586-021-03763-1} {\bibfield  {journal} {\bibinfo
  {journal} {Nature}\ }\textbf {\bibinfo {volume} {597}},\ \bibinfo {pages}
  {45} (\bibinfo {year} {2021})}\BibitemShut {NoStop}%
\bibitem [{\citenamefont {Davis}\ \emph {et~al.}(2023)\citenamefont {Davis},
  \citenamefont {Ye}, \citenamefont {Machado}, \citenamefont {Meynell},
  \citenamefont {Wu}, \citenamefont {Mittiga}, \citenamefont {Schenken},
  \citenamefont {Joos}, \citenamefont {Kobrin}, \citenamefont {Lyu},
  \citenamefont {Wang}, \citenamefont {Bluvstein}, \citenamefont {Choi},
  \citenamefont {Zu}, \citenamefont {Jayich},\ and\ \citenamefont
  {Yao}}]{davis2023}%
  \BibitemOpen
  \bibfield  {author} {\bibinfo {author} {\bibfnamefont {E.~J.}\ \bibnamefont
  {Davis}}, \bibinfo {author} {\bibfnamefont {B.}~\bibnamefont {Ye}}, \bibinfo
  {author} {\bibfnamefont {F.}~\bibnamefont {Machado}}, \bibinfo {author}
  {\bibfnamefont {S.~A.}\ \bibnamefont {Meynell}}, \bibinfo {author}
  {\bibfnamefont {W.}~\bibnamefont {Wu}}, \bibinfo {author} {\bibfnamefont
  {T.}~\bibnamefont {Mittiga}}, \bibinfo {author} {\bibfnamefont
  {W.}~\bibnamefont {Schenken}}, \bibinfo {author} {\bibfnamefont
  {M.}~\bibnamefont {Joos}}, \bibinfo {author} {\bibfnamefont {B.}~\bibnamefont
  {Kobrin}}, \bibinfo {author} {\bibfnamefont {Y.}~\bibnamefont {Lyu}},
  \bibinfo {author} {\bibfnamefont {Z.}~\bibnamefont {Wang}}, \bibinfo {author}
  {\bibfnamefont {D.}~\bibnamefont {Bluvstein}}, \bibinfo {author}
  {\bibfnamefont {S.}~\bibnamefont {Choi}}, \bibinfo {author} {\bibfnamefont
  {C.}~\bibnamefont {Zu}}, \bibinfo {author} {\bibfnamefont {A.~C.~B.}\
  \bibnamefont {Jayich}},\ and\ \bibinfo {author} {\bibfnamefont {N.~Y.}\
  \bibnamefont {Yao}},\ }\bibfield  {title} {\bibinfo {title} {Probing
  many-body dynamics in a two-dimensional dipolar spin ensemble},\ }\href
  {https://doi.org/10.1038/s41567-023-01944-5} {\bibfield  {journal} {\bibinfo
  {journal} {Nature Physics}\ }\textbf {\bibinfo {volume} {19}},\ \bibinfo
  {pages} {836} (\bibinfo {year} {2023})}\BibitemShut {NoStop}%
\bibitem [{\citenamefont {Martin}\ \emph {et~al.}(2023)\citenamefont {Martin},
  \citenamefont {Zhou}, \citenamefont {Leitao}, \citenamefont {Maskara},
  \citenamefont {Makarova}, \citenamefont {Gao}, \citenamefont {Zhu},
  \citenamefont {Park}, \citenamefont {Tyler}, \citenamefont {Park},
  \citenamefont {Choi},\ and\ \citenamefont {Lukin}}]{martin2023}%
  \BibitemOpen
  \bibfield  {author} {\bibinfo {author} {\bibfnamefont {L.~S.}\ \bibnamefont
  {Martin}}, \bibinfo {author} {\bibfnamefont {H.}~\bibnamefont {Zhou}},
  \bibinfo {author} {\bibfnamefont {N.~T.}\ \bibnamefont {Leitao}}, \bibinfo
  {author} {\bibfnamefont {N.}~\bibnamefont {Maskara}}, \bibinfo {author}
  {\bibfnamefont {O.}~\bibnamefont {Makarova}}, \bibinfo {author}
  {\bibfnamefont {H.}~\bibnamefont {Gao}}, \bibinfo {author} {\bibfnamefont
  {Q.-Z.}\ \bibnamefont {Zhu}}, \bibinfo {author} {\bibfnamefont
  {M.}~\bibnamefont {Park}}, \bibinfo {author} {\bibfnamefont {M.}~\bibnamefont
  {Tyler}}, \bibinfo {author} {\bibfnamefont {H.}~\bibnamefont {Park}},
  \bibinfo {author} {\bibfnamefont {S.}~\bibnamefont {Choi}},\ and\ \bibinfo
  {author} {\bibfnamefont {M.~D.}\ \bibnamefont {Lukin}},\ }\bibfield  {title}
  {\bibinfo {title} {Controlling {{Local Thermalization Dynamics}} in a
  {{Floquet-Engineered Dipolar Ensemble}}},\ }\href
  {https://doi.org/10.1103/PhysRevLett.130.210403} {\bibfield  {journal}
  {\bibinfo  {journal} {Physical Review Letters}\ }\textbf {\bibinfo {volume}
  {130}},\ \bibinfo {pages} {210403} (\bibinfo {year} {2023})}\BibitemShut
  {NoStop}%
\bibitem [{\citenamefont {He}\ \emph {et~al.}(2023)\citenamefont {He},
  \citenamefont {Ye}, \citenamefont {Gong}, \citenamefont {Liu}, \citenamefont
  {Murch}, \citenamefont {Yao},\ and\ \citenamefont {Zu}}]{he2023}%
  \BibitemOpen
  \bibfield  {author} {\bibinfo {author} {\bibfnamefont {G.}~\bibnamefont
  {He}}, \bibinfo {author} {\bibfnamefont {B.}~\bibnamefont {Ye}}, \bibinfo
  {author} {\bibfnamefont {R.}~\bibnamefont {Gong}}, \bibinfo {author}
  {\bibfnamefont {Z.}~\bibnamefont {Liu}}, \bibinfo {author} {\bibfnamefont
  {K.~W.}\ \bibnamefont {Murch}}, \bibinfo {author} {\bibfnamefont {N.~Y.}\
  \bibnamefont {Yao}},\ and\ \bibinfo {author} {\bibfnamefont {C.}~\bibnamefont
  {Zu}},\ }\bibfield  {title} {\bibinfo {title} {Quasi-{{Floquet
  Prethermalization}} in a {{Disordered Dipolar Spin Ensemble}} in
  {{Diamond}}},\ }\href {https://doi.org/10.1103/PhysRevLett.131.130401}
  {\bibfield  {journal} {\bibinfo  {journal} {Physical Review Letters}\
  }\textbf {\bibinfo {volume} {131}},\ \bibinfo {pages} {130401} (\bibinfo
  {year} {2023})}\BibitemShut {NoStop}%
\bibitem [{\citenamefont {Hughes}\ \emph {et~al.}(2025)\citenamefont {Hughes},
  \citenamefont {Meynell}, \citenamefont {Wu}, \citenamefont {Parthasarathy},
  \citenamefont {Chen}, \citenamefont {Zhang}, \citenamefont {Wang},
  \citenamefont {Davis}, \citenamefont {Mukherjee}, \citenamefont {Yao},\ and\
  \citenamefont {Jayich}}]{hughes2025}%
  \BibitemOpen
  \bibfield  {author} {\bibinfo {author} {\bibfnamefont {L.~B.}\ \bibnamefont
  {Hughes}}, \bibinfo {author} {\bibfnamefont {S.~A.}\ \bibnamefont {Meynell}},
  \bibinfo {author} {\bibfnamefont {W.}~\bibnamefont {Wu}}, \bibinfo {author}
  {\bibfnamefont {S.}~\bibnamefont {Parthasarathy}}, \bibinfo {author}
  {\bibfnamefont {L.}~\bibnamefont {Chen}}, \bibinfo {author} {\bibfnamefont
  {Z.}~\bibnamefont {Zhang}}, \bibinfo {author} {\bibfnamefont
  {Z.}~\bibnamefont {Wang}}, \bibinfo {author} {\bibfnamefont {E.~J.}\
  \bibnamefont {Davis}}, \bibinfo {author} {\bibfnamefont {K.}~\bibnamefont
  {Mukherjee}}, \bibinfo {author} {\bibfnamefont {N.~Y.}\ \bibnamefont {Yao}},\
  and\ \bibinfo {author} {\bibfnamefont {A.~C.~B.}\ \bibnamefont {Jayich}},\
  }\bibfield  {title} {\bibinfo {title} {Strongly {{Interacting}},
  {{Two-Dimensional}}, {{Dipolar Spin Ensembles}} in (111)-{{Oriented
  Diamond}}},\ }\href {https://doi.org/10.1103/PhysRevX.15.021035} {\bibfield
  {journal} {\bibinfo  {journal} {Physical Review X}\ }\textbf {\bibinfo
  {volume} {15}},\ \bibinfo {pages} {021035} (\bibinfo {year}
  {2025})}\BibitemShut {NoStop}%
\bibitem [{\citenamefont {Yoon}\ \emph {et~al.}(2026)\citenamefont {Yoon},
  \citenamefont {Oh}, \citenamefont {Lee},\ and\ \citenamefont
  {Choi}}]{yoon2026}%
  \BibitemOpen
  \bibfield  {author} {\bibinfo {author} {\bibfnamefont {T.}~\bibnamefont
  {Yoon}}, \bibinfo {author} {\bibfnamefont {S.}~\bibnamefont {Oh}}, \bibinfo
  {author} {\bibfnamefont {J.}~\bibnamefont {Lee}},\ and\ \bibinfo {author}
  {\bibfnamefont {H.}~\bibnamefont {Choi}},\ }\href
  {https://doi.org/10.48550/arXiv.2602.17074} {\bibinfo {title} {Mesoscopic
  {{Spin Coherence}} in a {{Disordered Dark Electron Spin Ensemble}}}}
  (\bibinfo {year} {2026}),\ \Eprint {https://arxiv.org/abs/2602.17074}
  {arXiv:2602.17074 [quant-ph]} \BibitemShut {NoStop}%
\bibitem [{\citenamefont {Yao}\ \emph {et~al.}(2014)\citenamefont {Yao},
  \citenamefont {Laumann}, \citenamefont {Gopalakrishnan}, \citenamefont
  {Knap}, \citenamefont {M{\"u}ller}, \citenamefont {Demler},\ and\
  \citenamefont {Lukin}}]{yao2014}%
  \BibitemOpen
  \bibfield  {author} {\bibinfo {author} {\bibfnamefont {N.~Y.}\ \bibnamefont
  {Yao}}, \bibinfo {author} {\bibfnamefont {C.~R.}\ \bibnamefont {Laumann}},
  \bibinfo {author} {\bibfnamefont {S.}~\bibnamefont {Gopalakrishnan}},
  \bibinfo {author} {\bibfnamefont {M.}~\bibnamefont {Knap}}, \bibinfo {author}
  {\bibfnamefont {M.}~\bibnamefont {M{\"u}ller}}, \bibinfo {author}
  {\bibfnamefont {E.~A.}\ \bibnamefont {Demler}},\ and\ \bibinfo {author}
  {\bibfnamefont {M.~D.}\ \bibnamefont {Lukin}},\ }\bibfield  {title} {\bibinfo
  {title} {Many-body localization in dipolar systems},\ }\href
  {https://doi.org/10.1103/PhysRevLett.113.243002} {\bibfield  {journal}
  {\bibinfo  {journal} {Physical Review Letters}\ }\textbf {\bibinfo {volume}
  {113}},\ \bibinfo {pages} {243002} (\bibinfo {year} {2014})}\BibitemShut
  {NoStop}%
\bibitem [{\citenamefont {Burin}(2015)}]{burin2015}%
  \BibitemOpen
  \bibfield  {author} {\bibinfo {author} {\bibfnamefont {A.~L.}\ \bibnamefont
  {Burin}},\ }\bibfield  {title} {\bibinfo {title} {Many-body delocalization in
  a strongly disordered system with long-range interactions: {{Finite}}-size
  scaling},\ }\href {https://doi.org/10.1103/PhysRevB.91.094202} {\bibfield
  {journal} {\bibinfo  {journal} {Physical Review B}\ }\textbf {\bibinfo
  {volume} {91}},\ \bibinfo {pages} {094202} (\bibinfo {year}
  {2015})}\BibitemShut {NoStop}%
\bibitem [{\citenamefont {Abanin}\ \emph {et~al.}(2019)\citenamefont {Abanin},
  \citenamefont {Altman}, \citenamefont {Bloch},\ and\ \citenamefont
  {Serbyn}}]{abanin2019}%
  \BibitemOpen
  \bibfield  {author} {\bibinfo {author} {\bibfnamefont {D.~A.}\ \bibnamefont
  {Abanin}}, \bibinfo {author} {\bibfnamefont {E.}~\bibnamefont {Altman}},
  \bibinfo {author} {\bibfnamefont {I.}~\bibnamefont {Bloch}},\ and\ \bibinfo
  {author} {\bibfnamefont {M.}~\bibnamefont {Serbyn}},\ }\bibfield  {title}
  {\bibinfo {title} {{\emph{Colloquium}} : {{Many}}-body localization,
  thermalization, and entanglement},\ }\href
  {https://doi.org/10.1103/RevModPhys.91.021001} {\bibfield  {journal}
  {\bibinfo  {journal} {Reviews of Modern Physics}\ }\textbf {\bibinfo {volume}
  {91}},\ \bibinfo {pages} {021001} (\bibinfo {year} {2019})}\BibitemShut
  {NoStop}%
\bibitem [{\citenamefont {Rezai}\ \emph {et~al.}(2025)\citenamefont {Rezai},
  \citenamefont {Choi}, \citenamefont {Lukin},\ and\ \citenamefont
  {Sushkov}}]{rezai2025}%
  \BibitemOpen
  \bibfield  {author} {\bibinfo {author} {\bibfnamefont {K.}~\bibnamefont
  {Rezai}}, \bibinfo {author} {\bibfnamefont {S.}~\bibnamefont {Choi}},
  \bibinfo {author} {\bibfnamefont {M.~D.}\ \bibnamefont {Lukin}},\ and\
  \bibinfo {author} {\bibfnamefont {A.~O.}\ \bibnamefont {Sushkov}},\
  }\bibfield  {title} {\bibinfo {title} {Probing dynamics of a two-dimensional
  dipolar spin ensemble using single qubit sensor},\ }\href
  {https://doi.org/10.1103/PhysRevLett.134.050801} {\bibfield  {journal}
  {\bibinfo  {journal} {Physical Review Letters}\ }\textbf {\bibinfo {volume}
  {134}},\ \bibinfo {pages} {50801} (\bibinfo {year} {2025})}\BibitemShut
  {NoStop}%
\bibitem [{\citenamefont {Van~Vleck}(1948)}]{vanvleck1948}%
  \BibitemOpen
  \bibfield  {author} {\bibinfo {author} {\bibfnamefont {J.~H.}\ \bibnamefont
  {Van~Vleck}},\ }\bibfield  {title} {\bibinfo {title} {The dipolar broadening
  of magnetic resonance lines in crystals},\ }\href
  {https://doi.org/10.1103/PhysRev.74.1168} {\bibfield  {journal} {\bibinfo
  {journal} {Physical Review}\ }\textbf {\bibinfo {volume} {74}},\ \bibinfo
  {pages} {1168} (\bibinfo {year} {1948})}\BibitemShut {NoStop}%
\bibitem [{\citenamefont {Bloembergen}(1949)}]{bloembergen1949}%
  \BibitemOpen
  \bibfield  {author} {\bibinfo {author} {\bibfnamefont {N.}~\bibnamefont
  {Bloembergen}},\ }\bibfield  {title} {\bibinfo {title} {On the interaction of
  nuclear spins in a crystalline lattice},\ }\href
  {https://doi.org/10.1016/0031-8914(49)90114-7} {\bibfield  {journal}
  {\bibinfo  {journal} {Physica}\ }\textbf {\bibinfo {volume} {15}},\ \bibinfo
  {pages} {386} (\bibinfo {year} {1949})}\BibitemShut {NoStop}%
\bibitem [{\citenamefont {Abragam}(2011)}]{abragam2011}%
  \BibitemOpen
  \bibfield  {author} {\bibinfo {author} {\bibfnamefont {A.}~\bibnamefont
  {Abragam}},\ }\href@noop {} {\emph {\bibinfo {title} {The Principles of
  Nuclear Magnetism}}},\ \bibinfo {edition} {repr}\ ed.,\ \bibinfo {series}
  {International Series of Monographs on Physics}\ No.~\bibinfo {number} {32}\
  (\bibinfo  {publisher} {Oxford Univ. Pr},\ \bibinfo {address} {Oxford},\
  \bibinfo {year} {2011})\BibitemShut {NoStop}%
\bibitem [{SM()}]{SM}%
  \BibitemOpen
  \href@noop {} {}\bibinfo {howpublished} {See Supplemental Material for
  derivations of the exact identities, Ising-field statistics, homogenization
  and P1 rate networks, surface exact diagonalization, and the blockade
  criterion in dense NV ensembles}\BibitemShut {NoStop}%
\bibitem [{\citenamefont {Anderson}(1958)}]{anderson1958}%
  \BibitemOpen
  \bibfield  {author} {\bibinfo {author} {\bibfnamefont {P.~W.}\ \bibnamefont
  {Anderson}},\ }\bibfield  {title} {\bibinfo {title} {Absence of diffusion in
  certain random lattices},\ }\href {https://doi.org/10.1103/PhysRev.109.1492}
  {\bibfield  {journal} {\bibinfo  {journal} {Physical Review}\ }\textbf
  {\bibinfo {volume} {109}},\ \bibinfo {pages} {1492} (\bibinfo {year}
  {1958})}\BibitemShut {NoStop}%
\bibitem [{\citenamefont {Klauder}\ and\ \citenamefont
  {Anderson}(1962)}]{klauder1962}%
  \BibitemOpen
  \bibfield  {author} {\bibinfo {author} {\bibfnamefont {J.~R.}\ \bibnamefont
  {Klauder}}\ and\ \bibinfo {author} {\bibfnamefont {P.~W.}\ \bibnamefont
  {Anderson}},\ }\bibfield  {title} {\bibinfo {title} {Spectral {{Diffusion
  Decay}} in {{Spin Resonance Experiments}}},\ }\href
  {https://doi.org/10.1103/PhysRev.125.912} {\bibfield  {journal} {\bibinfo
  {journal} {Physical Review}\ }\textbf {\bibinfo {volume} {125}},\ \bibinfo
  {pages} {912} (\bibinfo {year} {1962})}\BibitemShut {NoStop}%
\bibitem [{\citenamefont {Anderson}\ and\ \citenamefont
  {Weiss}(1953)}]{anderson1953}%
  \BibitemOpen
  \bibfield  {author} {\bibinfo {author} {\bibfnamefont {P.~W.}\ \bibnamefont
  {Anderson}}\ and\ \bibinfo {author} {\bibfnamefont {P.~R.}\ \bibnamefont
  {Weiss}},\ }\bibfield  {title} {\bibinfo {title} {Exchange narrowing in
  paramagnetic resonance},\ }\href {https://doi.org/10.1103/RevModPhys.25.269}
  {\bibfield  {journal} {\bibinfo  {journal} {Reviews of Modern Physics}\
  }\textbf {\bibinfo {volume} {25}},\ \bibinfo {pages} {269} (\bibinfo {year}
  {1953})}\BibitemShut {NoStop}%
\bibitem [{\citenamefont {Black}\ and\ \citenamefont
  {Halperin}(1977)}]{blackSpectralDiffusionPhonon1977}%
  \BibitemOpen
  \bibfield  {author} {\bibinfo {author} {\bibfnamefont {J.~L.}\ \bibnamefont
  {Black}}\ and\ \bibinfo {author} {\bibfnamefont {B.~I.}\ \bibnamefont
  {Halperin}},\ }\bibfield  {title} {\bibinfo {title} {Spectral diffusion,
  phonon echoes, and saturation recovery in glasses at low temperatures},\
  }\href {https://doi.org/10.1103/PhysRevB.16.2879} {\bibfield  {journal}
  {\bibinfo  {journal} {Physical Review B}\ }\textbf {\bibinfo {volume} {16}},\
  \bibinfo {pages} {2879} (\bibinfo {year} {1977})}\BibitemShut {NoStop}%
\bibitem [{\citenamefont {Lukin}\ \emph {et~al.}(2001)\citenamefont {Lukin},
  \citenamefont {Fleischhauer}, \citenamefont {Cote}, \citenamefont {Duan},
  \citenamefont {Jaksch}, \citenamefont {Cirac},\ and\ \citenamefont
  {Zoller}}]{lukin2001}%
  \BibitemOpen
  \bibfield  {author} {\bibinfo {author} {\bibfnamefont {M.~D.}\ \bibnamefont
  {Lukin}}, \bibinfo {author} {\bibfnamefont {M.}~\bibnamefont {Fleischhauer}},
  \bibinfo {author} {\bibfnamefont {R.}~\bibnamefont {Cote}}, \bibinfo {author}
  {\bibfnamefont {L.~M.}\ \bibnamefont {Duan}}, \bibinfo {author}
  {\bibfnamefont {D.}~\bibnamefont {Jaksch}}, \bibinfo {author} {\bibfnamefont
  {J.~I.}\ \bibnamefont {Cirac}},\ and\ \bibinfo {author} {\bibfnamefont
  {P.}~\bibnamefont {Zoller}},\ }\bibfield  {title} {\bibinfo {title} {Dipole
  {{Blockade}} and {{Quantum Information Processing}} in {{Mesoscopic Atomic
  Ensembles}}},\ }\href {https://doi.org/10.1103/PhysRevLett.87.037901}
  {\bibfield  {journal} {\bibinfo  {journal} {Physical Review Letters}\
  }\textbf {\bibinfo {volume} {87}},\ \bibinfo {pages} {037901} (\bibinfo
  {year} {2001})}\BibitemShut {NoStop}%
\bibitem [{\citenamefont {Bermudez}\ and\ \citenamefont
  {Schaetz}(2016)}]{bermudez2016}%
  \BibitemOpen
  \bibfield  {author} {\bibinfo {author} {\bibfnamefont {A.}~\bibnamefont
  {Bermudez}}\ and\ \bibinfo {author} {\bibfnamefont {T.}~\bibnamefont
  {Schaetz}},\ }\bibfield  {title} {\bibinfo {title} {Quantum {{Transport}} of
  {{Energy}} in {{Controlled Synthetic Quantum Magnets}}},\ }\href
  {https://doi.org/10.1088/1367-2630/18/8/083006} {\bibfield  {journal}
  {\bibinfo  {journal} {New Journal of Physics}\ }\textbf {\bibinfo {volume}
  {18}},\ \bibinfo {pages} {083006} (\bibinfo {year} {2016})},\ \Eprint
  {https://arxiv.org/abs/1512.03218} {arXiv:1512.03218 [quant-ph]} \BibitemShut
  {NoStop}%
\bibitem [{\citenamefont {De~Sousa}\ and\ \citenamefont
  {Das~Sarma}(2003)}]{desousaTheoryNuclearinducedSpectral2003}%
  \BibitemOpen
  \bibfield  {author} {\bibinfo {author} {\bibfnamefont {R.}~\bibnamefont
  {De~Sousa}}\ and\ \bibinfo {author} {\bibfnamefont {S.}~\bibnamefont
  {Das~Sarma}},\ }\bibfield  {title} {\bibinfo {title} {Theory of
  nuclear-induced spectral diffusion: {{Spin}} decoherence of phosphorus donors
  in {{Si}} and {{GaAs}} quantum dots},\ }\href
  {https://doi.org/10.1103/PhysRevB.68.115322} {\bibfield  {journal} {\bibinfo
  {journal} {Physical Review B}\ }\textbf {\bibinfo {volume} {68}},\ \bibinfo
  {pages} {115322} (\bibinfo {year} {2003})}\BibitemShut {NoStop}%
\bibitem [{\citenamefont {Zolotarev}(1986)}]{zolotarev1986}%
  \BibitemOpen
  \bibfield  {author} {\bibinfo {author} {\bibfnamefont {V.~M.}\ \bibnamefont
  {Zolotarev}},\ }\href@noop {} {\emph {\bibinfo {title} {One-Dimensional
  Stable Distributions}}},\ \bibinfo {series} {Translations of Mathematical
  Monographs}\ No.~\bibinfo {number} {65}\ (\bibinfo  {publisher} {American
  Mathematical Society},\ \bibinfo {address} {Providence, RI},\ \bibinfo {year}
  {1986})\BibitemShut {NoStop}%
\bibitem [{\citenamefont {Kittel}\ and\ \citenamefont
  {Abrahams}(1953)}]{kittel1953}%
  \BibitemOpen
  \bibfield  {author} {\bibinfo {author} {\bibfnamefont {C.}~\bibnamefont
  {Kittel}}\ and\ \bibinfo {author} {\bibfnamefont {E.}~\bibnamefont
  {Abrahams}},\ }\bibfield  {title} {\bibinfo {title} {Dipolar {{Broadening}}
  of {{Magnetic Resonance Lines}} in {{Magnetically Diluted Crystals}}},\
  }\href {https://doi.org/10.1103/PhysRev.90.238} {\bibfield  {journal}
  {\bibinfo  {journal} {Physical Review}\ }\textbf {\bibinfo {volume} {90}},\
  \bibinfo {pages} {238} (\bibinfo {year} {1953})}\BibitemShut {NoStop}%
\bibitem [{\citenamefont {Haan}\ and\ \citenamefont
  {Zwanzig}(1978)}]{haan1978}%
  \BibitemOpen
  \bibfield  {author} {\bibinfo {author} {\bibfnamefont {S.~W.}\ \bibnamefont
  {Haan}}\ and\ \bibinfo {author} {\bibfnamefont {R.}~\bibnamefont {Zwanzig}},\
  }\bibfield  {title} {\bibinfo {title} {F{\"o}rster migration of electronic
  excitation between randomly distributed molecules},\ }\href
  {https://doi.org/10.1063/1.435913} {\bibfield  {journal} {\bibinfo  {journal}
  {The Journal of Chemical Physics}\ }\textbf {\bibinfo {volume} {68}},\
  \bibinfo {pages} {1879} (\bibinfo {year} {1978})}\BibitemShut {NoStop}%
\bibitem [{\citenamefont {Gochanour}\ \emph {et~al.}(1979)\citenamefont
  {Gochanour}, \citenamefont {Andersen},\ and\ \citenamefont
  {Fayer}}]{gochanour1979}%
  \BibitemOpen
  \bibfield  {author} {\bibinfo {author} {\bibfnamefont {C.~R.}\ \bibnamefont
  {Gochanour}}, \bibinfo {author} {\bibfnamefont {H.~C.}\ \bibnamefont
  {Andersen}},\ and\ \bibinfo {author} {\bibfnamefont {M.~D.}\ \bibnamefont
  {Fayer}},\ }\bibfield  {title} {\bibinfo {title} {Electronic excited state
  transport in solution},\ }\href@noop {} {\bibfield  {journal} {\bibinfo
  {journal} {The Journal of Chemical Physics}\ }\textbf {\bibinfo {volume}
  {70}},\ \bibinfo {pages} {4254} (\bibinfo {year} {1979})}\BibitemShut
  {NoStop}%
\bibitem [{\citenamefont {Gr{\"a}{\ss}er}\ \emph {et~al.}(2023)\citenamefont
  {Gr{\"a}{\ss}er}, \citenamefont {Rezai}, \citenamefont {Sushkov},\ and\
  \citenamefont {Uhrig}}]{grasser2023}%
  \BibitemOpen
  \bibfield  {author} {\bibinfo {author} {\bibfnamefont {T.}~\bibnamefont
  {Gr{\"a}{\ss}er}}, \bibinfo {author} {\bibfnamefont {K.}~\bibnamefont
  {Rezai}}, \bibinfo {author} {\bibfnamefont {A.~O.}\ \bibnamefont {Sushkov}},\
  and\ \bibinfo {author} {\bibfnamefont {G.~S.}\ \bibnamefont {Uhrig}},\
  }\bibfield  {title} {\bibinfo {title} {Understanding the dynamics of randomly
  positioned dipolar spin ensembles},\ }\href
  {https://doi.org/10.1103/PhysRevResearch.5.043191} {\bibfield  {journal}
  {\bibinfo  {journal} {Phys. Rev. Research}\ }\textbf {\bibinfo {volume}
  {5}},\ \bibinfo {pages} {043191} (\bibinfo {year} {2023})}\BibitemShut
  {NoStop}%
\bibitem [{\citenamefont {Choi}\ \emph {et~al.}(2020)\citenamefont {Choi},
  \citenamefont {Zhou}, \citenamefont {Knowles}, \citenamefont {Landig},
  \citenamefont {Choi},\ and\ \citenamefont {Lukin}}]{choi2020prx}%
  \BibitemOpen
  \bibfield  {author} {\bibinfo {author} {\bibfnamefont {J.}~\bibnamefont
  {Choi}}, \bibinfo {author} {\bibfnamefont {H.}~\bibnamefont {Zhou}}, \bibinfo
  {author} {\bibfnamefont {H.~S.}\ \bibnamefont {Knowles}}, \bibinfo {author}
  {\bibfnamefont {R.}~\bibnamefont {Landig}}, \bibinfo {author} {\bibfnamefont
  {S.}~\bibnamefont {Choi}},\ and\ \bibinfo {author} {\bibfnamefont {M.~D.}\
  \bibnamefont {Lukin}},\ }\bibfield  {title} {\bibinfo {title} {Robust
  {{Dynamic Hamiltonian Engineering}} of {{Many-Body Spin Systems}}},\ }\href
  {https://doi.org/10.1103/PhysRevX.10.031002} {\bibfield  {journal} {\bibinfo
  {journal} {Physical Review X}\ }\textbf {\bibinfo {volume} {10}},\ \bibinfo
  {pages} {031002} (\bibinfo {year} {2020})}\BibitemShut {NoStop}%
\end{thebibliography}%
	
	\section*{End Matter}
	\section*{Appendix A: Stable law and the regime number}
	For Poisson-distributed points, $\langle\prod_lf(\mathbf r_l)\rangle=\exp[-\rho\int d^dr\,(1-f)]$. For $S^z_l=\pm\frac12$ with equal weights, $f=\cos[kQ(\mathbf r)/2]$. A bath polarization adds only an imaginary part to the exponent, so the width does not depend on it~\cite{SM}. With $s(\Omega)=J_0|a(\Omega)||k|/2$,
	\begin{equation}
		-\ln\langle e^{ikh_j}\rangle=\rho\int d\Omega\int_0^\infty r^{d-1}dr\,\big[1-\cos(s/r^3)\big].
		\label{eq:A0}
	\end{equation}
	The substitution $u=s/r^3$ and the integral $\int_0^\infty u^{-1-\mu}(1-\cos u)\,du=\Gamma(1-\mu)\cos(\pi\mu/2)/\mu$ at $\mu=d/3$ give Eq.~\eqref{eq:stable} with $K_d=\Gamma(1-d/3)\cos(\pi d/6)/d$, i.e., $K_3=\pi/6$ (as a limit) and $K_2=\Gamma(\frac13)/4$. In three dimensions, $\int|a|\,d\Omega=16\pi/3\sqrt3$ leads to Eq.~\eqref{eq:GammaL}. At 110~ppm this gives $\Gamma_L/2\pi=2.55$~MHz, while lattice Monte Carlo sampling gives 2.54--2.61~MHz. In two dimensions, with $a=-\cos2\phi$, one has $\int_0^{2\pi}|\cos2\phi|^{2/3}d\phi=2\sqrt\pi\,\Gamma(\frac56)/\Gamma(\frac43)$.
	
	A symmetric stable law with exponent $C|k|^\alpha$ has the density $\Gamma(1+1/\alpha)/(\pi C^{1/\alpha})$ at zero. For the pair, $C=2c_h$, so that $P_{ij}(0)=\Gamma(1+3/d)/[\pi(2c_h)^{3/d}]$. The strongest coupling of a spin to partners of density $\nu\rho$, with $|J(\mathbf r)|=J_0|a|/4r^3$, follows $\Pr(|J_1|<x)=\exp[-\nu\rho V_d(x)]$ with $V_d(x)=d^{-1}(J_0/4x)^{d/3}\int|a|^{d/3}d\Omega$. Its median is $|J_1|_{1/2}=\frac{J_0}{4}\big[\nu\rho\int|a|^{d/3}d\Omega/(d\ln2)\big]^{3/d}$, and in the product the density and the angular integral cancel,
	\begin{equation}
		2\pi|J_1|_{1/2}P_{ij}(0)=\frac4\lambda\,\Gamma\Big(1+\frac3d\Big)\Big[\frac{\nu}{2dK_d\ln2}\Big]^{3/d},
		\label{eq:A1}
	\end{equation}
	which is Eq.~\eqref{eq:regime}. The factor $4/\lambda$ appears because the Ising field at fixed $J_{ij}$ scales as $\lambda$. The fraction of spins whose strongest bond exceeds the golden-rule bound is $1-\exp\{-\ln2\,[\pi|J_1|_{1/2}P_{ij}(0)]^{d/3}\}$, which is 0.25 on the surface and 0.013--0.052 for the P1 groups. Monte Carlo sampling of Poisson configurations reproduces Eq.~\eqref{eq:stable} and the medians $|J_1|_{1/2}$ within the sampling error~\cite{SM}.
	
	\section*{Appendix B: Spin lock and excluded channels}
	With $J_{ij}=-Q_{ij}/4$, one has $H_{\rm dd}=\sum_{i<j}Q_{ij}[S^z_iS^z_j-\frac12(S^x_iS^x_j+S^y_iS^y_j)]$. We add the lock $\Omega\sum_kS^y_k$ and on-site fields $\sum_j\delta_jS^z_j$. In the toggling frame of the lock, averaging over one period gives
	\begin{multline}
		\bar H=-\frac12\sum_{i<j}Q_{ij}\Big[S^y_iS^y_j-\frac12\big(S^z_iS^z_j+S^x_iS^x_j\big)\Big]\\
		+\sum_j\frac{\delta_j^2}{2\Omega}S^y_j+O\Big(\frac{\|H_{\rm dd}\|^2}{\Omega}\Big).
		\label{eq:B1}
	\end{multline}
	The dipolar part is secular about $y$ with $Q'_{ij}=-Q_{ij}/2$ and $J'_{ij}=-J_{ij}/2$. Hence $\lambda'=4$, and the dressed resonance window of a bond is $|\eta'_{ij}|<2|J'_{ij}|=|J_{ij}|$. A global rotation by $\pi/2$ about $x$ maps $\bar H|_{\delta=0}$ onto $-\frac12H_{\rm dd}$. Since infinite-temperature autocorrelations are even in time, $\langle S^y_j(t)S^y_j\rangle_{\bar H}=C_j(t/2)$, where $C_j$ is the autocorrelation of $S^z_j$ under $H_{\rm dd}$, and therefore $T_{1\rho}/2=T_z|_{\delta\equiv0}+O(\Omega^{-1})$. For a random six-spin cluster at $\Omega=400\max|Q_{ij}|$, the two sides agree to four decimal places.
	
	At the lowest drive in the plateau, $\Omega=79~\mu$s$^{-1}$, the proton field and the NV field enter as $\delta_j^2/2\Omega\le0.12$ and $\le0.15~\mu$s$^{-1}$, respectively, below the dressed window $|J_{ij}|\simeq0.9~\mu$s$^{-1}$. Static Larmor offsets below $2\pi\times1.9$~MHz fall inside this window, and larger offsets would make the rate depend on $\Omega$. Noise-induced flips scale as $\Omega^{-2}$ ($5\times10^{-4}~\mu$s$^{-1}$), higher-order dipolar terms as $|J_{ij}|^2/\Omega$ ($\approx0.01~\mu$s$^{-1}$), and drive inhomogeneity as $\Omega$. Each of these contributions would change by at least a factor of 5.3 across a plateau that is constant within $\pm20\%$. The only other $\Omega$-independent channel is lattice relaxation. If $T_1$ is a property of the defect type, then $T_1\ge57~\mu$s, which contributes at most $0.018~\mu$s$^{-1}$ to the plateau rates $1/T_{1\rho}=0.057$--$0.104~\mu$s$^{-1}$.
	
	\section*{Appendix C: Global control cannot tune $\lambda$}
	For like spin-$\frac12$ dipoles, each coupling can be written as $\sum_{\alpha\beta}S^\alpha_iT^{\alpha\beta}_{ij}S^\beta_j$ with $T_{ij}=Q_{ij}\,{\rm diag}(-\frac12,-\frac12,1)$, which is symmetric and traceless. A global rotation $R$ maps $T\to R^{\rm T}TR$, and averaging over any sequence preserves both properties. U(1) symmetry about an axis $\mathbf m$ requires $\bar T=a\,\mathbf m\mathbf m^{\rm T}+b(I-\mathbf m\mathbf m^{\rm T})$, and tracelessness gives $a+2b=0$. In the form $QS^zS^z+J(S^+S^-+S^-S^+)$, one has $Q=a$ and $J=b/2$, hence $Q=-4J$ ($\lambda=4$) or $a=b=0$.
	
	\begin{table}[b]
		\caption{\label{tab:inputs}Inputs and results. Top: P1 bath~\cite{zu2021}, with $\rho=0.0194$~nm$^{-3}$, $J_0/2\pi=52.0$~MHz\,nm$^3$, $\gamma=0.5~\mu$s$^{-1}$ (S2: 0.3), $x_s=0.1$~MHz, $\Gamma_L/2\pi=2.55$~MHz, and $c=1.94$; $D$ in nm$^2/\mu$s; ``window'': networks fitted over 30--300~$\mu$s. Bottom: the four spin-lock systems of Ref.~\cite{rezai2025}, digitized from Fig.~4(c) of that work to within $\lesssim5\%$ and ordered by increasing proton-field amplitude; $S$ from Eq.~\eqref{eq:S}.}
		\begin{ruledtabular}
			\begin{tabular}{lcccc}
				P1 condition & network & window & no Ising & measured $gD$\\
				\hline
				$\nu=\frac13$, S1 & 0.76 & $0.82\pm0.11$ & 9.0 & $0.82\pm0.17$\\
				$\nu=\frac13$, S2 & 0.71 & $0.78\pm0.11$ & 9.3 & $1.03\pm0.13$\\
				$\nu=\frac14$ & 0.57 & $0.66\pm0.09$ & 10.9 & $0.74\pm0.18$\\
				$\nu=\frac14$, driven & 0.66 & $0.75\pm0.10$ & 10.9 & $0.95\pm0.26$\\
				$\nu=\frac1{12}$ & 0.124 & $0.19\pm0.04$ & 2.5 & $0.33\pm0.08$\\
				\hline
				Surface system & $T_2$ ($\mu$s) & $T_{1\rho}/2$ ($\mu$s) & $T_{1\rho}/2T_2$ & $S$\\
				\hline
				1 & 0.63 & 8.7 & 13.9 & 8.5\\
				2 & 0.65 & 8.7 & 13.3 & 8.1\\
				3 & 0.65 & 4.8 & 7.4 & 4.5\\
				4 & 0.84 & 5.7 & 6.8 & 4.1\\
			\end{tabular}
		\end{ruledtabular}
	\end{table}
\end{document}